\documentclass[preprint]{aastex}

\shorttitle{Close binary stars.~XV}
\shortauthors{Pribulla \& et al.}

\begin{document}

\title{Radial Velocity Studies of Close Binary
Stars.~XV\footnote{Based on the data obtained at the David Dunlap
Observatory, University of Toronto.}}

\author{
Theodor Pribulla\altaffilmark{1}, Slavek M. Rucinski,
R.M. Blake\altaffilmark{2}, Wenxian Lu\altaffilmark{3},
J.R. Thomson, Heide DeBond, Toomas Karmo, Archie de Ridder}
\affil{Department of Astronomy \& Astrophysics, University of Toronto \\
50 St.\ George Street, Toronto, ON  M5S~3H4}
\email{pribulla@ta3.sk,(rucinski,debond,karmo,ridder)@astro.utoronto.ca}
\author{Waldemar Og{\l}oza, Greg Stachowski}
\affil{Mt.\ Suhora Observatory of the Pedagogical University\\
ul.~Podchora\.{z}ych 2, 30--084 Cracow, Poland}
\email{(ogloza,greg)@ap.krakow.pl}
\author{Michal Siwak}
\affil{Astronomical Observatory, Jagiellonian University, ul. Orla 171,
       30--244 Cracow, Poland}
\email{siwak@oa.uj.edu.pl}

\altaffiltext{1}
{Astronomical Institute, Slovak Academy of Sciences,
059 60 Tatransk\'a Lomnica, Slovakia}

\altaffiltext{2}
{Current address: Dept. of Physics and Earth Science, University of North Alabama,
 Florence, 35632 AL, U.S.A., email: rmblake@una.edu
}

\altaffiltext{3}
{Current address: Department of Geography and Meteorology and
 Department of Physics and Astronomy, Valparaiso University, Valparaiso,
 IN 46383, U.S.A., e-mail: Wen.Lu@valpo.edu}

\begin{abstract}
Radial-velocity measurements and sine-curve fits to the orbital radial
velocity variations are presented for the last
eight close binary systems analyzed the same way as
in the previous papers of this series: QX~And,
DY~Cet, MR~Del, HI~Dra, DD~Mon, V868~Mon, ER~Ori, and Y Sex. For another
seven systems (TT~Cet, AA~Cet, CW~Lyn, V563~Lyr, CW~Sge, LV~Vir and MW~Vir)
phase coverage is insufficient to provide reliable
orbits but radial velocities of individual components were
measured. Observations of a few complicated systems
observed throughout the DDO close-binary
program are also presented; among them an
especially interesting is the multiple system
V857~Her which -- in addition to the
contact binary -- very probably contains one or more sub-dwarf
components of much earlier spectral type. All suspected binaries which
were found to be most probably pulsating
stars are briefly discussed in terms of mean radial velocities
and projected rotation velocities ($v \sin i$) as well as
spectral type estimates. In two of
them, CU~CVn and V752~Mon, the broadening functions show
a clear presence of non-radial pulsations.
The previously missing spectral types for the DDO~I paper
are given here in addition to such estimates for most of the
program stars of this paper.
\end{abstract}


\section{INTRODUCTION}
\label{sec1}

This is the last in a series of papers presenting results of
spectroscopic observations taken within the program of
radial velocity (hereafter RV) orbits of close binary stars
at the David Dunlap Observatory (DDO); it contains a discussion of
the all remaining targets of this program.
For full references to the previous papers, see the last paper by
\citet[Paper XIV]{ddo14}; for technical details and conventions, for
preliminary estimates of uncertainties, and for a description of the
broadening functions (BFs) technique, see the interim summary paper
\citet[Paper VII]{ddo7}.

Most of data used in the present paper were obtained -- as in most of the
previous fourteen papers -- using the broadening
function approach applied to the
spectral region of the Mg~I triplet (always centered at 5184~\AA).
Up to August 2005 we used a diffraction grating with 1800 lines/mm, after which
 a new one with 2160 lines/mm was used. A small number of
observations were taken in the red region centered at 6290\AA\
which includes a telluric band to resolve our concerns on the
stability of the RV system
(see \citet[Paper XI]{ddo11}). The RV observations reported in this paper
have been collected between October 1996 and July 2, 2008, the day
when the David Dunlap Observatory ceased to operate. The ranges
of dates for individual systems can be found in Table~\ref{tab1}.
We note that this program utilized the efficient code
of \citet{pych2004} for removal of cosmic rays from 2-D images.

Throughout our program, selection of the targets was quasi-random:
At a given time, we observed a few dozen
close binary systems with periods usually
shorter than one day, brighter than 10 -- 11 magnitude and with
declinations $>-20^\circ$; we published the results in groups of ten systems
as soon as reasonable orbital elements were obtained from measurements evenly
distributed in orbital phases. This paper is an exception as we have
not been able to collect ten orbits before the observatory closure and
we were left with material for a few stars already started.
For that reason, in
this last paper of the series, we present all remaining spectroscopic
observations. Thus, they include: (i)~radial velocity orbits for
eight close binaries observed and fully analyzed
in the same way as in previous publications of this series
(QX~And, DY~Cet, MR~Del, HI~Dra, DD~Mon, V868~Mon, ER~Ori, and Y~Sex);
(ii)~close binaries with just one quadrature covered or with a few
spectra available but not sufficient in number to define a reasonable orbit
(TT~Cet, AA~Cet, CW~Lyn, V563~Lyr, CW~Sge, LV~Vir, and MW~Vir);
(iii)~complicated or faint binaries/triples not providing reliable
RVs (GO~Cyg, V857~Her, V752~Mon, V353~Peg, and MS~Vir),
(iv)~additional data and improved orbits
for a few multiple systems already presented
within this series (ET~Boo, XY~Leo, and TV~UMi);
(v)~radial velocities and projected
rotational velocities, $v \sin i$, for several stars found to be
pulsating rather than being close binaries,
and (vi)~the previously missing
spectral-type estimates for close binaries in \citet[Paper I]{ddo1}.
All the RVs are given in Table~\ref{tab1}.
Among the present targets, only QX~And, ER~Ori, and Y~Sex,
have had spectroscopic orbits previously published while
preliminary orbits of AA~Cet and TT~Cet, based on limited data,
were given by \citet{south2}.

The RVs for the eight
short period binaries reported in this paper were determined
by fitting the double rotational profiles to extracted BFs, as explained in
\citet{ddo11}. Similarly as in our previous papers dealing with multiple systems
(here the cases of MR~Del, V563~Lyr, DD~Mon, ER~Ori, Y~Sex, and LV~Vir),
the RVs for the
eclipsing pair were obtained after removal of the slowly rotating component,
as was described most recently in \citet{ddo14}.

As in other papers of this series, whenever possible, we estimated spectral
types of the program stars using new classification spectra centered
at 4200 \AA\ or 4400 \AA.
Additional classification spectra were obtained for part of
the systems published in \citet[Paper I]{ddo1} where no classifications
were given. The estimated spectral types were compared
with the mean $(B-V)$ color indices usually taken from the Tycho-2 catalog
\citep{Tycho2} and the photometric estimates of the spectral types using the
relations of \citet{Bessell1979}. In this paper we also made use of infrared
colors determined from the 2$\mu$ All Sky Survey (2MASS, \citet{2MASS}).
Especially useful is the $J-K$ color index,
which is monotonically rising from the early
spectral types to about M0V \citep{cox2000}; this index is relatively
less affected by the interstellar absorption than $B-V$.
Parallaxes cited throughout the paper were adopted from the new reduction
of the Hipparcos raw data \citep{newhipp} which supersedes the original
reductions \citep{hip}.

The RV data for the binaries
are given in Table~\ref{tab1}. The preliminary
sine-curve solutions for the eight binaries are in Table~\ref{tab2}
while the phase diagrams are shown in Figures~\ref{fig1} and \ref{fig2}.
Section~\ref{sec2} of the paper contains
summaries of previous studies for individual systems with reliable orbits and
comments on the new data. Systems with the
insufficient orbit coverage are discussed
in Section~\ref{sec3}, while problematic systems for which RVs
could not be determined are discussed in Section~\ref{sec4}.
Targets which were found to be pulsating
are presented in Section~\ref{sec5}. Examples of BFs of
individual  systems are shown in Figures~\ref{fig3} -- \ref{fig5}.

The data for the eight close binaries
in Table~\ref{tab2} are organized in the same manner as in
the previous papers of this series.
In addition to the parameters of spectroscopic orbits,
the table provides information about the relation between the
spectroscopically observed upper conjunction of the more massive
component, $T_0$ (not necessarily the primary eclipse)
and the recent photometric determinations of the primary
minimum in the form of the $O-C$ deviations for the number of
elapsed periods $E$. The reference ephemeris of
HI~Dra was taken from \citet{gome1996} and for V868~Mon from
\citet{oter2004}. For the rest of systems the ephemerides were adopted from
the on-line version of ``An Atlas of O-C diagrams of eclipsing binary
stars''\footnote{http://www.as.wsp.krakow.pl/ephem/} \citep{Kreiner2004}.
Because the on-line ephemerides are frequently updated, we
give those used for the computation of the $O-C$ residuals below
Table~\ref{tab2} (the status as of October 2008).
The deeper eclipse in W-type
contact binary systems corresponds to the lower conjunction of the more
massive component; in such cases the epoch in Table~\ref{tab2} is a
half-integer number.

\section{EIGHT STARS WITH RELIABLE ORBITS}
\label{sec2}

\subsection{QX~And}

QX~And (GSC~2816~1950, H235 in NGC~752) is a contact
binary star in the intermediate age
open cluster NGC~752. Its variability was first
noted by \citet{john1953} on the
basis of two discordant photometric observations.
\citet{plat1991} gave 99\% probability of the cluster
membership for QX~And based on the proper motion. The
first thorough photometric and spectroscopic observations of QX~And
were presented by \citet{milo1995}.
The authors found that QX~And is a contact binary, estimated its
spectral type as F3-5, determined absolute parameters of the components
(masses $M_1 = 1.18\, M_\odot$ and $M_2 = 0.24\, M_\odot$)
by simultaneous fits to the observed light and
RV variations and found the distance to the
binary and the cluster as $381 \pm 17$ pc.
The center-of-mass velocity $V_0 = +11.7 \pm 2.7$ km~s$^{-1}$
(given in their Table~16), was larger than the cluster
mean velocity, $RV = +5.5$ km~s$^{-1}$ \citep{dani1994}.
Unfortunately, the RV observations
of \citet{milo1995} were of a rather poor quality (see their Fig.~5)
resulting in only marginally useful orbital parameters.

Preliminary results of the DDO spectroscopy of
QX~And were published as a part
of the PhD thesis of \citet{blak2002}. Here we present an independent
determination using the rotational-profile fitting to the
extracted BFs. Our spectroscopic orbit (Table~\ref{tab2}) is not fully
consistent with the previous result of \citet{milo1995}. In particular,
the systemic velocity, $V_0 = +4.1 \pm 1.2$ km~s$^{-1}$ is now
close to that of the cluster, while both, the
mass ratio $q = 0.306 \pm 0.009$ and the projected total mass,
$(M_1 + M_2) \sin^3 i = 1.038 \pm 0.022\, M_\odot$ are larger.

Although QX~And was not observed by the Hipparcos
satellite due to its low brightness, an independent estimate of
its distance can be obtained using the absolute magnitude calibration
of \citet{rd1997}. Using the photometry of \citet{milo1995} giving
$V_{max} = 11.49$ and the de-reddened color $(B-V)_0 = 0.43$ (corresponding
to the F5 spectral type), we obtain $M_V = 3.13$ and the distance modulus,
$(V - M_V) = 8.36$ (470 pc), which is close to the range
considered for NGC~752 by \citet{schi1988}:
$(V - M_V) = 8.17 \pm 0.15$ from a full binary solution for
DS~And and $7.9 \pm 0.2$ from isochrone fitting to
un-evolved Main Sequence stars.

\subsection{DY~Cet}

Variability of DY~Cet (HIP~12311, HD~16515) was detected during the Hipparcos
mission where it was classified as a W~UMa binary. The system has been
neglected by observers in spite of its large
photometric amplitude, $\Delta H_p = 0.56$, possibly
because of the negative declination ($-$14\degr 18\arcmin).
A preliminary analysis of the
Hipparcos light curve (hereafter LC) was
performed by \citet{sela2004}, who -- using the
Fourier-coefficient method -- determined the
following geometric elements: mass ratio
$q_{ph} = 0.45$, fill-out $f = 0.2$, and inclination angle $i = 77.5\degr$.
Except for a few times of minima, \citep{kraj2006,dwor2006,kraj2007},
no systematic, ground-based observations of the system have been performed.

Because of the low sky elevation as seen from DDO, our
spectroscopic observations of DY~Cet were obtained over a long period
of time.
The orbital solution (Table~\ref{tab2}, Fig.~\ref{fig1}) is based on
two combined datasets, one using the 1800 lines/mm grating and a THX
chip and the second, more recent, obtained with the
2160 lines/mm grating and a JY2 chip. Because of the long time interval
over which the spectra were taken, it was necessary to optimize the
period length, in addition to the initial phase. The RVs
show that the deeper minimum is an upper conjunction of the
less massive component so that the system is of the A subtype.

The Hipparcos parallax of DY~Cet, $\pi = 5.11 \pm 1.99$ mas, is of
limited value for a luminosity estimate. The
Tycho-2 color index, $(B-V) = 0.39$, corresponds to a
F3/4V spectral type, which would be expected
for a contact binary with an orbital period of 0.44 days. The
2MASS infrared color, $(J-K) = 0.258$, is also consistent with
a spectral type of F4/5V. Unfortunately, we have not obtained
any classification spectra for the star.

\subsection{MR~Del}

MR~Del (HIP~101236, HD195434) is a part of a visual binary composed
of $V = 9.49$ and $V = 9.77$ components, currently separated by 1.8\arcsec\,
at a position angle of 71\degr\, \citep{wds}. \citet{cuti1997} found that
the brighter visual component is an active eclipsing binary with
$P = 0.52175(22)$ days and inferred spectral type of K for all three
components. The out-of-eclipse photometric wave with an amplitude
of about 0.04 mag was interpreted by the authors as due to photospheric
spots. The large proper motion of the whole triple system,
0.416\arcsec /year, when combined with its low metallicity,
$\log {\rm [m/H]} = -0.99$ (relative to the Sun)
indicates that it is a halo or old disk object.
Photometric observations of MR~Del were later obtained by \citet{clau2001}
(in the $uvby$ system) and by \citet{soyd2001}.
No spectroscopy of the system has yet been available.

Our spectra included all components of the visual pair. The third component
with $L_3/(L_1 + L_2) = 0.51 \pm 0.06$ (at the maximum light of the
eclipsing binary) showed a stable radial velocity, $V_3  = -51.95 \pm 0.35$
km~s$^{-1}$, close to the systemic velocity of the close binary
$V_0  = -49.8 \pm 0.8$ km~s$^{-1}$. The broadening functions
suggest that the eclipsing binary is a close, detached system,
but without any indications of photospheric
spots previously implied by the LC asymmetry and related to the
strong X-ray emission \citep{cuti1997}.
With the new, reliably determined mass ratio and the projected total
mass $(M_1 + M_2) \sin^3 i$,
the system requires a new photometric analysis to determine the
inclination angle and thus its absolute parameters.

The Hipparcos parallax, $20.72 \pm 2.49$ mas, although of low accuracy,
can be combined with the proper motion to give the space
velocity of the system, $V = 107 \pm 10$ km~s$^{-1}$.
Unfortunately, we have no classification spectra for the
components of the system.

\subsection{HI~Dra}

The variability of HI~Dra (HIP~90972, HD~171848) was
detected during the Hipparcos mission where it was
classified as a RR~Lyr pulsating star. HI~Dra
was later photometrically observed by \citet{gome1999}, who suggested
that it is a $\beta$~Lyrae or an ellipsoidal binary system. When folding
their new data and the Hipparcos photometric observations
with the period of 0.597417 days, the authors noticed an
O'Connell effect of 0.02 mag which supported
the binary interpretation.
The  Hipparcos photometry was re-analyzed by \citet{sela2004}
using the Fourier-coefficient method. The author found
that the system is very probably a contact binary and obtained the first set
of parameters: the mass ratio $q = 0.15$, inclination
angle $i = 52.5\degr$ and fill-out $f = 0.7$.

Our spectroscopy and the appearance of the BFs (Fig.~\ref{fig3})
confirm that HI~Dra is a contact binary star.
The resulting mass ratio, $q = 0.25$, is substantially
larger than that found by \citet{sela2004} documenting the
unreliability of photometric mass ratios for partially
eclipsing binaries and especially systems with low
photometric amplitudes. It is rather surprising, however,
that the time of the lower conjunction of the more massive
component occurs in photometric phase zero which indicates
a W-subtype; for a low mass ratio, one would expect
the A-subtype. But it is possible that the orbital period
is changing or that the initial epoch was not reliably
determined from the Hipparcos photometry

The catalog spectral type of HI~Dra, F8, does not
correspond to the observed $J-K = 0.123$ \citep{2MASS}
nor to $B-V = 0.272$ \citep{Tycho2}.
The relatively long orbital period of HI~Dra also
suggests an earlier spectral type. Although we do not have
any classification spectra, we found that the
best BF template for the observations
around the 5184\AA\ triplet was for the spectral type
F0-F2V. The Hipparcos parallax of the system is rather small,
$3.93 \pm 0.68$ mas.

\subsection{DD~Mon}

The variability of DD~Mon (HD~292319, GSC~4800~372)
was first noticed by \citet{hoff1934}.
The system was studied thoroughly by \citet{yama1990}, who obtained
BV photometric data and medium dispersion (37\AA/mm) photographic spectra.
Analysis of their LCs showed that DD~Mon is a near-contact or a semi-detached
binary with its primary component almost filling its Roche lobe. Due to the low
brightness of the system only the primary component was identified in the
spectra resulting in: $V_0 = 8.1 \pm 1.5$ km~s$^{-1}$, $K_1 = 89.1 \pm 2.2$
km~s$^{-1}$. The spectral type was estimated as F5IV/V. A LC
analysis led to the mass ratio $q = 0.70 \pm 0.15$ and the inclination
$i = 80 \pm 0.4\degr$. Later, \cite{qian1997} obtained new $BV$ photometry
of the system and contrary to \citet{yama1990} found that the secondary
component fills its Roche lobe.

The system is rather faint, m$_{pg}$ = 11.10 at the light maximum, and we
were able to take only a few spectra of it.
They show three components: In addition to
the rapidly rotating components of the eclipsing pair,
the BFs clearly show a slowly-rotating third
component at $V_3 = 19.5 \pm 2.6$ km~s$^{-1}$,
with $L_3/(L_1+L_2) \simeq 0.22$.
Our orbit is based on only 9 RV measurements for the primary and 8 RVs of
the secondary component. Therefore we regard the resulting parameters as
very preliminary. The binary system mean velocity,
$V_0 = 20.2 \pm 1.9$ km~s$^{-1}$,
is equal within its error to that of the third component
indicating a physical bond. DD~Mon was not known to be a multiple system
before and it is not listed in the WDS Catalogue \citep{wds}.
The profiles of the close-binary components have the same projected
rotation velocity $v \sin i \simeq 130$ km~s$^{-1}$, with the secondary
significantly fainter (Fig.~\ref{fig3}), indicating detached or semi-detached
configuration.

We have no spectral classification spectra for DD~Mon.
The spectral type of the system is probably rather late in spite of the
original classification in the HD Catalogue of B5.
The Tycho-2 color $B-V = 0.510$
indicates the spectral type F8 while the
2MASS infrared color $J-K = 0.388$ suggests a spectral type
as late as G4. The best fitting template for the DD~Mon spectra was
HD~187691 which has a spectral type F8V.

\subsection{V868~Mon}

V868~Mon (BD$-2~2221$, GSC~4835~1947) is a relatively
bright ($V = 8.9$) W~UMa-type eclipsing
binary with photometric amplitude of about 0.50 mag.
It was discovered on Stardial images
by \citet{wils2003}. The authors noted that the
eclipses are probably total and determined
the following ephemeris for the primary minimum
HJD $2452681.731 + 0.63772 \times E$. Later
\citet{oter2004} identified this variable star in the NSVS
photometry\footnote{http://www.skydot.lanl.gov/nsvs/nsvs.php}
and improved its orbital period to $P=0.637705$ days.
The authors classified this binary as EB.
No further observations or investigation of the system
have been published since then.

Our spectroscopy shows that V868~Mon is indeed a
very close or contact binary; the profiles
of the components are never completely separated.
The RVs of the secondary component are
rather uncertain especially around the second orbital quadrature, where the
profiles usually appear to be deformed or asymmetric,
indicating an intrinsic variability. An orbital period of \citet{oter2004}
was adopted in our RV orbit solution; our
attempts to adjust the period
did not lead to any significant improvement of the solution.

V868~Mon has a rather high systemic velocity, $V_0 = 78.1$ km~s$^{-1}$.
Unfortunately, the system was not included
in the Hipparcos mission so that its distance is not known
and the proper motion components are not
precise enough to determine its space velocity.
The 2MASS infrared color index of V868~Mon, $(J-K) = 0.16$,
corresponds to a F0 spectral type while the
Tycho-2 $(B-V)$ = 0.20 to about A8V.
The best template to the spectra,
HD~128167, has a spectral type of F2V, suggesting a
slightly later spectral type.


\subsection{ER~Ori}

The variability of ER~Ori (HIP~24156, WDS~05112--0833)
was first noticed by \citet{hoff1929} while the
variability type was properly assigned by \citet{flor1931}.
Later, \citet{stru1944} obtained a
preliminary spectroscopic orbit of the system which is,
however, of marginal use because of the inadequate spectral
resolution and a small number of observations.
Because of its large photometric
amplitude ($\Delta V \sim 0.6$) and short period (0.4234 days),
the system was subject to many photometric studies.
The photometric analyzes lead to a rather
inconsistent set of geometric parameters;
for references and discussion see \citet{kim2003}. A wave-like
pattern in the time-of-minima (O-C) diagram was
interpreted in terms of an invisible body on a 35-years orbit by
\citet{abhy1982}.

The breakthrough investigation in the study of ER~Ori
was that of \citet{goec1994}. The authors
found that (i)~the system is totally eclipsing with a
totality lasting 12.6 minutes, (ii)~a third
component appears in the cross-correlation functions of the spectra
at a place established using speckle interferometry data
(separation $\rho = 0.187(7)\arcsec$,
position angle $\theta = 354.1(15)\degr$,
intensity ratio $L_3/(L_1+L_2) = 0.16(3)$ in
the $H_\alpha$ filter, as of March 1993),
(iii)~the orbital period of the third body, as
estimated from the binary period variations, is about
63 years and the orbit is seen almost pole-on,
(iv)~the first, reliable
spectroscopic orbit based on photographic spectra with parameters:
$K_1 = 130.1 \pm 6.8$ km~s$^{-1}$,
$K_2 = 235.5 \pm 6.9$ km~s$^{-1}$,
$V_0 = 37.9 \pm 3.3$ km~s$^{-1}$; with the photometrically
determined inclination angle $i = 87.5\degr$, the masses are:
$M_1 = 1.39 \pm 0.10\, M_\odot$ and $M_2 = 0.76 \pm 0.08\, M_\odot$;
(vi)~the third component RV is
constant, $V_3 = 41 \pm 1.5$ km~s$^{-1}$,
and within the error of the systemic velocity of the close binary.

Adaptive optics and infrared observations of \citet{ruci2007}
confirmed the presence of the third component. In
January 1998, it was practically at the same position ($\rho = 0.183\arcsec$,
$\theta = 354.4\degr$, $\Delta K = 2.14$) as found
by \citet{goec1994}, while in October 2005 it was not detectable,
indicating it was closer than the detection limit of
about $0.09\arcsec$ (for the observed $\Delta K$).
The disappearance of the companion almost coincides
with its predicted periastron passage (July/August 2004)
\citep{kim2003}, as based on the observed light-time
effect (hereafter LITE). The periastron passage in 2004, however,
is not fully supported by the large acceleration terms
observed during the Hipparcos program\footnote{In the new
reductions of \citet{newhipp}, the astrometric solution did not
require any acceleration terms.},
$g_\alpha = -19.26 \pm 6.57$ mas~yr$^{-2}$
and $g_\delta = -17.34 \pm 4.71$ mas~yr$^{-2}$.

Our spectroscopic elements are different from
those of \citet{goec1994}:
$K_1 = 148.0 \pm 2.3$ km~s$^{-1}$, $K_2 = 225.5 \pm 2.3$ km~s$^{-1}$,
$V_0 = 49.4 \pm 1.7$ km~s$^{-1}$; the mass ratio is slightly
larger, $q = 0.656(12)$.
The velocity of the third component was found to vary significantly
between +38 km~s$^{-1}$ (HJD 2\,454\,180) and +20
km~s$^{-1}$ (HJD 2\,454\,530).
This means that the third component itself is very probably
also a binary. Another possibility, that the
third component is a single star and revolves around the contact pair,
cannot be ruled out completely as our present, good quality data
still do not permit to detect the motion of the third
component reflected in the contact-binary RVs. The
light contribution of the third component,
as found from the integrated BF's,
is $\beta = L_3/(L_1 + L_2) =  0.16 \pm 0.02$, which is
fairly consistent with the result of \citet{goec1994}.

Although ER~Ori was observed by the Hipparcos satellite,
its parallax, $\pi = 4.47 \pm 3.50$ mas \citep{newhipp},
is basically useless (we note that it was strongly negative in the
original Hipparcos results). The parallax problems are --
very probably -- due to the presence of
the visual companion 13.6\arcsec\,  to the north and to the
orbital motion in the tight visual pair.
The Hipparcos astrometry could possibly be re-analyzed
utilizing observed LITE and a distance estimate from
the absolute-magnitude calibration
of \citet{rd1997}. Our own spectral type estimate is F7/8.
Assuming F8 on the basis of $(B-V)_0 = 0.54$,
we obtain $M_V = 3.41$. Taking $V_{max} = 9.325$ \citep{goec1994}
and $\beta = 0.16$, we have $V_{max}^{12} = 9.49$
and then $d \simeq 164$ pc or $\pi \simeq 6$ mas.


\subsection{Y~Sex}

Y~Sex (HIP~49217) is a contact binary within
a tight visual double (WDS 10028+0106,
HDS~1451, $\rho = 0.49\arcsec$,
$\theta = 154\degr$, $V = 10.08 + 12.70$).
The variability of Y~Sex was first noted by
\citet{hoff1934} and the first
photoelectric photometry of the system was obtained
by \citet{tana1957}. Later \citet{hill1979} analyzed its
LC assuming the Roche model. He
found it to be of the W-subtype and showing total eclipses; his
determination of the geometric elements gave:
$1/q = 5.72 \pm 0.07$, $i = 76.8\degr$
and a marginal contact $f = 0$. \citet{lean1983} obtained the
first spectroscopic observations of the system. By measuring the
metal line centroids those authors obtained
the following orbital elements:
$V_0  = 9.8 \pm 6$ km~s$^{-1}$, $K_1 = 40.0 \pm 8$
km~s$^{-1}$ and $K_2 = 218 \pm 31$ km~s$^{-1}$ resulting in the mass ratio,
$q = 0.18 \pm 0.03$, i.e.\ compatible with the photometric determination
of \citet{hill1979},
which is not surprising since the system is totally eclipsing.

Later studies of Y~Sex
\citep{herz1993,qian2000,wolf2000,heqi2007} concentrated
on the variable orbital period of the system. \citet{wolf2000} found that
the sinusoidal variation of the orbital period can be interpreted by the
presence of an invisible third body on a 58-years orbit and estimated its
spectral type as M4/5 and $M_3 \simeq 0.3\,M_\odot$. Finally, \citet{yang2003}
detected a tiny third light $L_3 = 0.0064 \pm 0.0008$ in their LC solution.
It is surprising that all investigators somehow overlooked the fact that
Y~Sex is part of a tight visual binary detected by the Hipparcos
satellite \citep{hip}.

Y~Sex was somewhat faint for the DDO instrumentation which together with its
relatively early spectral type led to noisy broadening functions.
The very close companion separated by $0.49\arcsec$ was always within the
seeing disk at DDO. Its light contribution
$L_3/(L_1+L_2) =  0.11 \pm 0.03$, as determined from the BFs close to
the orbital quadratures of the eclipsing pair,
is slightly larger than that inferred from the magnitude
difference as cataloged in the WDS.

The 2MASS $J-K = 0.278$ index is consistent with the
F5 classification while Tycho-2 $B-V = 0.39$ corresponds
to a slightly earlier spectral type F3/4V.
The Hipparcos parallax, $\pi = 8.75 \pm 2.09$
mas is rather imprecise to be of any use.
Our classification spectra indicate the spectral type F5/6.

\section{BINARY STARS WITH INSUFFICIENT ORBITAL COVERAGE}
\label{sec3}

Several targets were observed at DDO but a sufficient phase coverage was
not achieved either due to low brightness (resulting in poor
spectra), low sky elevation or
the DDO closure before the conclusion of this program.
Below, we describe systems for which RV data were collected,
but usually only one orbital quadrature was adequately covered. Such data
do not define the mass ratio reliably but can be used to estimate
the total projected semi-amplitude $(K_1 + K_2)$ and thus the
total projected mass, or can be combined with future observations.

\subsection{TT~Cet}

Properties of TT~Cet (HIP~8294) are discussed and a
preliminary orbit given in \citet{south2}.
It is a close, but very probably a detached or semi-detached binary.
The system was found to be too faint and too low over the southern,
city-illuminated sky
for DDO ($\delta = -9\degr 45\arcmin$, $V_{max}$ = 10.9).
The quality of the extracted BFs was further deteriorated
by the fairly early spectral type and thus
a correspondingly weak Mg~I triplet.

Most of BFs extracted from the DDO spectra show only the primary component;
the secondary is visible in only one BF, see Fig.~\ref{fig3}. Our RVs
(Table~\ref{tab1}) are given only
for the primary component. The 2MASS color index $J-K = 0.25$
indicates the F4/5V spectral type while the Tycho-2 $B-V=0.404$
corresponds to a F3/4 spectral type. Our own spectral
classification confirms a spectral type of F4V.

\subsection{AA~Cet}

For a fuller discussion of this system, see \citet{south2}.
Because AA~Cet (HIP~9258, HD~12180) was visible very low from DDO, we
were able to take only 11 spectra of this system; unfortunately,
most of them happened to be close to the conjunctions. Therefore, we
are giving only a few RV
determinations for AA~Cet in Table~\ref{tab1}. The extracted BFs show a
systematic difference between the quadratures: While around the phase 0.25,
the profile of the secondary is close to the
expected rotational profile, the profiles for the phases around 0.75
appear narrower and triangular or irregular shape.
Our classification spectra indicate the spectral type F4V.

\subsection{CW~Lyn}

Variability of CW~Lyn (HIP 42554) was discovered by the Hipparcos
mission, where it was classified as a $\beta$~Lyrae variable with
0.812389 days orbital period. Later \citet{sela2004} analyzed the
Hipparcos LC using the Fourier-coefficient method and
found that the system is a genuine contact binary. The author
determined the following preliminary parameters: a fill-out $f = 0.0$,
mass ratio $q = 0.10$, inclination angle $i = 77.5\degr$. The
Hipparcos LC with the amplitude of about $\Delta V = 0.25$
indicates a possibility of total eclipses for a low mass ratio
system. No ground-based observations of the system have been published
so far.

Our spectroscopy covers the second orbital quadrature only. The Hipparcos
ephemeris does not satisfactorily predict
phases of our spectroscopy indicating
either a significant period change or a problem with the ephemeris.
The BFs (see Fig.~\ref{fig3}) support the low mass ratio of the system but
show some peculiarities similar to those
seen in AW~UMa \citep{prib2008}: (i)~the secondary
has a triangular shape,  (ii)~the primary component has a wide base underneath
a relatively narrow profile, in a clear disagreement with the Roche model.

The system was included in the Hipparcos mission, but its
parallax, $\pi  = 3.93 \pm 1.37$ mas, is of limited use. The
BFs strengths indicate the F3/4 spectral type which is fairly consistent with
the 2MASS color $J-K = 0.263$. Our classification spectrum
corresponds to the same spectral type, F4V.

\subsection{V563~Lyr}

Variability of V563~Lyr (NSV~11321) was noted by \citet{hoff1966} and
the system was later photometrically observed by \citet{belt1999}, who
found it to be a contact binary with $P = 0.577639$ days, and estimated its
spectral type as F5. The LC minima seemed to be of similar depth.
The authors also noted that the components are very probably evolved
because of the rather late spectral type for the orbital period.
Except for a few minima observations, the system has never been studied in detail.

With a visual magnitude ranging between 10.96 and 11.47 \citep{belt1999},
the system was difficult for the DDO spectral observations.
Most of the spectra were noisy so that
only 8 observations during two nights of excellent seeing
were useful for RV determinations
(see Table~\ref{tab1}). The BFs immediately show that
the system is accompanied by a third component with a light contribution
of about $L_3/(L_1+L_2)$ = 0.15. Unfortunately, only one orbital
quadrature was covered with good observations
preventing a reliable determination of the
orbit. Assuming that the third component is physically bound, and that
its velocity, $V_3 \sim = 14$ km~s$^{-1}$, is identical to the
center-of-mass velocity of the contact binary,
we can roughly estimate the mass ratio at $q = 0.37$
and the total projected mass at a relatively high
value of $(M_1 + M_2) \sin^3 i = 3.0\,M_\odot$.

The trigonometric parallax of V563~Lyr is unknown since
it was not included in the Hipparcos mission.
Its 2MASS infrared color $J-K = 0.216$
corresponds to the F2/3 spectral type while the Tycho-2
color is $B-V = 0.456$,
implying F5V is probably affected by the interstellar extinction.

\subsection{CW~Sge}

CW~Sge (HIP 98430) is a rather faint ($V_{max} = 11.0$)
contact binary. Its variability was noticed by \citet{hoff1935}
who classified the system as a RR~Lyr
variable with $P = 0.330223$ days.
Later \citet{lang1960} found it to be a W~UMa variable with a
two times longer period.
In the catalogue of \citet{bran1980}, CW~Sge is a close but detached
binary with components filling respectively
86\% and 85\% of their Roche radii.
No other photometric or spectroscopic study of
the system has yet been published
. The adaptive optics observations of \citet{ruci2007}
showed that the binary is accompanied by
a late-type companion separated by 1.84\arcsec.
This companion most probably entered the slit of the DDO spectrograph
during periods of bad seeing. CW~Sge may in fact host
an even closer companion as indicated by a
large, 7.64 mas, ``cosmic error'' in the original Hipparcos astrometric
solution \citep{hip}, leading to a negative parallax,
$\pi = -0.23 \pm 2.41$ mas.

CW~Sge was found to be too faint for DDO and was abandoned.
Only one orbital quadrature (phase 0.75)
was observed making a reliable orbit determination
impossible. The best template for the observed spectra is that of a
F8 spectral type, which appears to be too late for a contact binary with
orbital period as long as $P = 0.66$ days. The RVs show that the more
massive component is the cooler one. Our own spectral classification
spectrum suggests a spectral type of F6, which is moderately
consistent with $J-K=0.23$ which suggests type F3.
%

\subsection{LV~Vir}

Variability of LV~Vir (HIP~66078, HD~117780)
was detected by the Hipparcos mission,
where it was classified as a $\beta$~Lyrae variable
with orbital period $P = 0.409439$ days.
LV~Vir is member of the relatively close
visual binary WDS~13328--1746 consisting of
$V=9.09$ and $V=9.43$ components, currently at the
position angle $\theta = 27\degr$ and the separation $\rho=1.2\arcsec$.
This resulted in the inclusion of
both components into the DDO spectrograph slit.
The Fourier analysis of the Hipparcos LC by
\citet{sela2004} showed that LV~Vir is a genuine contact binary.
Except for speckle interferometry observations of
the visual pair, no ground-based photometry or spectroscopy of
the system has been published yet, in spite of the
fairly large photometric amplitude of the variable
($\Delta V = 0.20$).

Because of the southern declination of the star, we were able to
take only 3 spectra of the system. The resulting BFs (see
Fig.~\ref{fig3}) show all three components (the binary and the visual
companion) and support the contact configuration for the
close binary system. Although we do not have a classification spectrum
for the system, the best match to the observed spectra of LV~Vir
indicates the F6/7 spectral type. The 2MASS infrared
color $J-K = 0.282$ corresponds to the F5 spectral type
while the Tycho-2 $B-V=0.521$ corresponds to the
F7/8 spectral type.

\subsection{MW~Vir}

MW~Vir (HIP~69828, HD~125048) was found to be variable during
the Hipparcos mission. The wavelike variation with
$\Delta H_p = 0.03$ and $P = 0.246539$ days could either
result from pulsations of a single star or from ellipsoidal variations
in a close binary. Although the system is a very bright one,
$V \simeq 7.0$, no ground-based observations of the star have been
published yet. MW~Vir was identified
with an X-ray source both in the EINSTEIN and ROSAT observations
indicating a late-type
companion to the moderately early-type star \citep{chis1999}.

Nine spectra taken close to the conjunction show
that MW~Vir is definitely a binary
star. Unfortunately, only one component is spectrally
visible with determinable
RV so that the secondary is probably a low mass,
late-type star. The rotational velocity of the primary is about
$v \sin i \simeq 75$ km~s$^{-1}$; its orbital RV's were observed
to decrease by 33 km~s$^{-1}$ over the
phase interval of 0.123 (see Table~\ref{tab1}).

The system is relatively nearby, with the Hipparcos parallax of
$12.40 \pm 1.10$ mas. The Tycho-2 $B-V = 0.229$ corresponds to
a spectral type of A7-8V, while the 2MASS $J-K = 0.186$ gives
a substantially later type, F1V. Assuming that the system
is a contact binary and using the absolute magnitude
calibration of \citet{rd1997}
we find $M_V^{cal} = 2.18$ (for $B-V = 0.23$) which is consistent with the
absolute magnitude given by the Hipparcos parallax, $M_V = 2.42$.

\section{DIFFICULT BINARY STARS}
\label{sec4}

In this section we present fragmentary results for binary
systems where -- although
attempted -- no reliable measurements of
RVs could be obtained. Although only indicative, these
results may lead to further research as some
of the objects appear to be very interesting.

The problems that we
encountered were as follows: (i)~the
target was observed too low on the DDO sky resulting in poor
spectra; (ii)~the spectral type was found to be rather early
making the lines of the Mg~I triplet too weak; (iii)~the target
was too faint for our telescope -- spectrograph combination;
(iv)~the close binary was found to be
accompanied by a third, bright and dominant,
rapidly-rotating companion spectrally obscuring
the binary; (v)~the third component had a significantly
different spectral type from the binary and suppressed
the binary signature in the BFs.

\subsection{GO~Cyg}

GO~Cyg (HIP~101748, HD~196628) is a
close but very probably detached eclipsing binary system;
for references and details see \citet{seze1993,oh2000}. Its
spectroscopic orbit based on RVs obtained by the CCF method was
presented in \citet{seze1993}.

The system consists of components of different temperatures
($T_1$ = 10,700 K and $T_2$ = 6,200 K) and thus it is not ideal for
the BF approach; moreover, the
primary is too early to have sufficiently strong
lines in the Mg~I region. Only 4 spectra of the system
were taken, all close to the secondary minimum. In addition to
the primary component of the binary,
one can clearly see an additional component rotating with
$v \sin i \sim = 35$ km~s$^{-1}$ (see Fig.~\ref{fig4}).
It is interesting to note that the third component
was not noted in the CCFs analysis
presented by \citet{seze1993}. A reliable analysis of the system would
require a high-dispersion spectroscopy over the whole
optical range for a spectral disentanglement of all components.

\subsection{V857~Her}

The variability of V857~Her (GSC~3070~345, $V_{max} = 10$, sp.\ type A6)
was noticed by \citet{gey1955}, who proposed it to be a Cepheid pulsating
star. No systematic observations of the system were performed
until 1996, when \citet{gome1996} obtained extensive CCD photometry
of the system and found it to be a totally eclipsing W~UMa binary with
$P = 0.3825$ days. A preliminary solution of the LCs --
mostly driven by the small photometric amplitude and the
presence of total eclipses -- led to
a very low mass ratio, $q_{ph} = 0.0725 \pm 0.050$, and a
high fill-out factor $f = 0.80 \pm 0.10$. The authors also noticed
cyclic LC changes. Later, \citet{qian2005}
determined an even smaller (in fact, a record small)
mass ratio, $q_{ph} = 0.0653$. No spectroscopic
study of the system has been published so far.

Two sets of spectra of V857~Her were obtained at DDO, the older set using
the 1800 lines/mm grating and THX chip and the more recent set
using the 2160 lines/mm grating
and the JY2 chip. The spectral lines in Mg~I region are
surprisingly shallow
resulting in poor BFs for either of the DDO datasets; basically, no
trustworthy RVs could be determined. Averaging of
all available spectra for either of the datasets has
resulted in reasonable mean BF
when analyzed with a F2 template (Fig.~\ref{fig4});
the BFs corresponded to a single, relatively rapidly-rotating star with
a presumed presence of another component in the system. The secondary
component was not visible due to a (possibly) low mass ratio and --
obviously -- to the averaging of the spectra.
Subtracting the corresponding binary contribution (obtained by a
convolution of the template spectrum with the BF) results in a
residual spectrum resembling that of a late B-type or an early
A-type star rotating at $v \sin i \simeq 50$ km~s$^{-1}$.
Because this approach involves an average spectrum, we cannot
say if the early-type component detected that way
is a binary or a single star. If this
component is physically connected with the close binary,
it must be a hot subdwarf, either a single or binary star.
V857~Her did not appear to be a visual double on the
spectrograph slit and is not listed in the WDS catalogue \citep{wds}.

The peculiarity of the system was noticed
before by \citet{priruc2006} in that it appeared too
blue for its orbital period of 0.3825 d.
The Tycho-2 color index $(B-V) = 0.164$,
corresponds to the A6 spectral type
while our classification spectrum indicates
an A7 spectral type. The presence of an early-type component would
explain the period -- color discrepancy and the shallowness of
the eclipses but would also put in
doubt the very low mass ratio previously indicated by the
LC modeling. V857~Her is a very interesting system and certainly
deserves a dedicated study based on high S/N, wide
wavelength coverage (echelle) spectra.

\subsection{V752~Mon}
\label{v752mon}

V752~Mon (HIP~34401, HD~54250) is member of
the visual binary WDS~07079--0441 ($V = 7.47 + 7.95$,
$\theta = 24\degr$, $\rho = 1.7\arcsec$). During the Hipparcos mission
it was found that the object is a low-amplitude variable and was
tentatively classified as a $\delta$~Scuti pulsator. The
rather early spectral type, A9V -- F0V, supported
this classification. However, knowing how many Hipparcos
short-period pulsating stars turned out to be close binaries, we
included V752~Mon in our binary star program.

Unfortunately, the orientation of the visual pair was not
convenient so that on most nights both components were
simultaneously included in the spectrograph slit.
The brighter component
dominated in the spectra, but the extracted BFs do not show any changes
when arranged versus the presumed pulsation period or
versus a period two times longer. In fact,
the BFs show only one rapidly rotating component with
$v \sin i \sim 155-160$ km~s$^{-1}$, with no traces of the
visual companion which would be expected for two almost
identical components with only a 0.5 magnitude
difference \citep{wds}.

The dominating star in the spectra appears to
be a $\delta$~Scuti star. The BFs clearly show wavelike structures
propagating on the surface of the rapidly rotating
star which can be explained by non-radial modes
with $m \simeq 8 - 10$.
We describe a similar case with very clear signatures of
non-radial pulsations, CU~CVn, in Section~\ref{sec5}
(see Fig.~\ref{fig5}) where we comment on the applicability
of the BF technique to pulsating stars.

At this point, we cannot exclude a possibility that the
photometric (binary star) variability comes from the secondary component
of the visual pair and that pulsations of the dominating star simply
mask the spectral signatures of a close binary in that star.
It is puzzling that the visual secondary
is not present in our spectra at all.  Its
absence may be explained in two ways: (1)~it is also
a rapidly rotating star at the (almost) same RV
as the primary so that the two broad profiles overlap, or
(2)~it is in fact a close binary, but because our spectra -- in the
mediocre seeing conditions -- were guided on the brighter
component of the pair, the light contribution of the secondary
to the spectra was lesser than expected. Obviously, only spectra
of individual components of the visual pair will establish
which component is variable and what is its type.

We note that
if we were sure that a close binary is not hiding as the secondary
of the visual pair, we would have
included V752~Mon in Section~\ref{sec5}, among pulsating stars
serendipitously observed by us during our program.

\subsection{V353~Peg}

V353~Peg (HIP~116108) is a bright ($V_{max}=7.42$)
low-amplitude ($\Delta H_p = 0.07$)
variable discovered by the Hipparcos satellite and
classified as a $\beta$~Lyrae-type eclipsing binary with a period
of  0.584557 days. The eclipsing pair is very probably accompanied by a
third component because the system shows a non-linear sky motion
as indicated by its astrometric solution requiring an
acceleration term (the ``G'' flag in H59)\footnote{In the new reduction
of Hipparcos raw data \citep{newhipp},
there is ``stochastic'' solution for V353~Peg instead}.
Unfortunately, speckle interferometry of \citet{maso2001}
could not resolve the visual pair. V353~Peg
was observed spectroscopically as part of the ground-based spectroscopic
support of the Hipparcos mission \citep{fehr1997}: the four RV observations
showed a spread of 40 km~s$^{-1}$. Otherwise, this bright variable
has been rather neglected.

The BFs show a strong, dominating, slowly-rotating component
accompanied by very faint features of a close binary.
The BFs extracted around the binary quadratures
conclusively show the phase variations expected for a close binary.
However, the binary is just
barely distinguishable at the base of the strong peak of the
companion (Fig.~\ref{fig4}) so its RV's could not be well
determined. The BF analysis was further
complicated by a significant difference in the spectral
types of the binary and its bright companion
resulting in a depression around the dominant component which
affects the residual profile of the binary.
The third component is significantly brighter than
the eclipsing binary with $L_3/(L_1+L_2) \simeq 2.5$ preventing
a sound analysis of the system. The
RV of the third component was found to be practically
constant at $V_3 \simeq -5$ km~s$^{-1}$ (see Table~\ref{tab3}).

The 2MASS infrared color $J-K = 0.158$ corresponds to
the F0V spectral type which is fairly consistent with
the Tycho-2 $B-V=0.21$ implying an A8 type. Our own,
rather uncertain classification
indicates a combination of spectral types, A7 and F2,
possibly corresponding to the dominant third component
and the eclipsing binary.

\subsection{MS~Vir}

MS~Vir (HIP~68881) is another Hipparcos discovery. The
system was originally classified as a $\beta$~Lyrae
eclipsing binary with
the variability period of $P = 0.31244$ days.
According to \citet{sela2004}, who modeled
the Hipparcos photometry using the Fourier-coefficient method, MS~Vir is
not a $\beta$~Lyr system but a contact binary with the geometric
parameters: $q = 0.25$, $f = 1.00$, and $i = 52.5\degr$.

Because the system was visible
very low at DDO ($\delta = -17\degr 41\arcmin$), we obtained only
7 spectra of the system. Although we attempted to obtain well exposed
spectra using 900 sec exposures,
the spectra and resulting BFs are too poor
to enable reliable RV measurements.
The BFs indicate that MS~Vir is indeed a close binary
composed of similar components.
New spectroscopy from the Southern hemisphere is
needed to derive reliable parameters of the system. We note that the
Hipparcos parallax of the star is rather large,
$\pi = 13.36 \pm 1.65$ mas, so that the system is nearby.

\section{LOW AMPLITUDE AND PULSATING VARIABLES}
\label{sec5}

\subsection{Low amplitude variables}

The Hipparcos mission have detected many low amplitude variables
with variability periods between 0.1 -- 0.5 days. As the
original analyzes and subsequent studies
have shown, the variables turned out to be mostly pulsating stars of
$\delta$~Sct, RR~Lyr, SX~Phe and $\gamma$~Dor types with an admixture of
contact binaries seen at low inclination angles. As was demonstrated by
\citet{Rci2001}, the probability of very small
amplitudes for contact binaries
does not go to zero with $i \rightarrow 0$ but rather stabilizes
at a finite value. Additionally,
low amplitudes of some among the close binaries
could be caused by the presence of a third component.
In such cases, companions may force bluer colors of the combined
systems making them more similar to pulsating stars; this would invalidate
the period-color diagram approach of \citet{duer1997}
of finding contact binaries among small amplitude variables.
Our series and its companion investigations on triplicity of
close binaries \citep{priruc2006,dangelo2006,ruci2007}
have shown that the frequency of companions
to close binaries is exceptionally high and may be approaching
100\%. For that reason, we felt that many stars
classified as pulsating by the Hipparcos project
required checking for the possibility of being in fact binaries.
In most cases, we stopped further observing
after establishing that a star is a sharp-lined one and thus
not a close binary. But, sometimes, when the
presence of a dominating third component was suspected, we continued
our observations. Thus, in this somewhat erratic way, we collected
several RV observations which may be of use in the future.

Throughout the DDO close binary program
we stumbled upon 17 low-amplitude
variables which were hard to classify and where only one
spectral component seen at a constant RV was detected.
For these systems we give mean RV, the projected
rotational velocity, $v \sin i$ and
estimated spectral type in Table~\ref{tab5}.
The individual RVs and rotational velocities
are given in Table~\ref{tab6}.
Of course, we do not exclude a possibility that some of the
constant velocity stars are in fact close binaries, possibly with
companions. Here, we simply report our findings for any future use.

\subsection{V752 Mon and CU CVn}

The most obvious radially pulsating stars observed in this program
are V752~Mon (this system may contain a close binary, Section~\ref{v752mon})
and CU~CVn. Both show rapid rotation and in both, both ``bumps'' or ``ripples''
are clearly present in their BFs when these
are arranged in phase into two-dimensional displays
(Fig.~\ref{fig5}). In both cases,
the ripples move in the direction of the stellar rotation and are
uniformly distributed along the stellar longitudes clearly indicating
non-radial pulsations. Such ripples have been observed before
in individual lines or in cross-correlation analysis studies
of $\delta$~Sct stars
\citep{mante97,mante04}. Here, we see them very well defined in
the broadening functions which attests the usefulness of this
technique even for pulsating stars: Apparently, all lines in
our spectral window of some 240 \AA\ behaved and traced
the pulsations the same way.

We have more data for V752~Mon than for CU~CVn
(Fig.~\ref{fig5}), but the quality
is poorer, possibly because of a more complex nature of the multiple
system and the possible presence of the close binary
(Section~\ref{v752mon}).
It is very interesting that
we can distinguish the very stable, practically
totally unmodified ripple pattern after one year (exactly
342 days) separating the first and the last nights of the V752~Mon
observations. The non-radial modes appear to be
prograde ones with $m \simeq 8 - 10$.

In the case of CU~CVn the ripples are even better defined and
can be easily measured for the temporal period of return of
$P \sim  0.4-0.5$ days. \citet{vidal2002} rejected the original
Hipparcos classification of the star as a W~UMa binary (which was
the  reason for us to observe the star) and gave the dominant
pulsation period as 0.0678 days, i.e.\ one half of the
Hipparcos period.  The appearance of the ripples, compared with
theoretical simulations \citep{koch2004},
indicate sectoral mode with $l = m = 6 - 8$ and
an axial inclination angle $i \sim 90\degr$.

Both, V752~Mon and CU~CVn, definitely deserve new
high-resolution spectroscopy. We suspect a similar picture to
emerge for IN~Dra (see below)
but -- with only a few spectra for this star --
we cannot state anything definitely.

\subsection{Other single or pulsating stars}

The main spectroscopic results for low amplitude variables
observed during the
DDO binary program are given in Table~\ref{tab5}.
Unambiguous variability classification have been possible only for a
few of them because of the small number of spectra or short, one-night
durations of some of our observations. These limitations apply especially
to HV~Eri, V1359~Ori and GG~UMa, for which we cannot reject a
possibility that they are eclipsing variables seen at low
inclination angle; these systems could either be SB1 or
observed close to either of the conjunctions.

In the pulsating star CC~Lyn, the DDO data probably
relate only to the brighter component of the visual pair.
The RV of CC~Lyn was found to vary between
+15 km~s$^{-1}$ and +23 km~s$^{-1}$.

\section{SPECTRAL CLASSIFICATIONS}
\label{sec6}

During our program we realized the value
of independently acquired spectral
classifications. For that reason, we planned to classify all binaries
without available spectral types.
Unfortunately, we have not been able to achieve this for
all binaries of this series before the observatory closure.
In addition to the data for the eight binaries which form the main results
of this paper, as given in Table~\ref{tab2}, we obtained
classification spectra only for systems
originally studied in the first paper of
the series \citet[Paper I]{ddo1}. The spectral types were
determined for: V417~Aql (F7V), LS~Del (F5-8V),
EF~Dra (F8/9V), V829~Her (F6-9V),
UV~Lyn (G0V), BB~Peg (F7V), and AQ~Psc (F5-8V). Also, we determined
the spectral types for three stars which were
later found to be pulsating: FH~Cam (A8V),
CU~CVn (A7V), and CC~Lyn (F0-2V).

\section{QUADRUPLE SYSTEMS}

Four quadruple systems were previously
observed and analyzed at DDO: ET~Boo, XY~Leo, VW~LMi, and TV~UMi
\citep{ddo11,ddo12}.
Because of the high significance of these results for our
understanding of the stability of those tight quadruple systems and to
improve the orbits of the companion (not close) binaries, we continued
acquisition of the spectra for these systems until the closure
of the observatory.

New RV observations of VW~LMi
were published in a subsequent dedicated paper \citep{vwlmi},
while the data for the remaining systems are listed in Table~\ref{tab3}.
The corresponding orbital elements for ET~Boo, XY~Leo,
and TV~UMi are given in Table~\ref{tab4}. For TV~UMi,
the new observations enabled us to determine the
orbital period and the spectroscopic elements; we note that
both were indeterminable in the original
paper \citep{ddo11} of the DDO series. The
RVs of TV~UMi and their best fits are shown in Fig.~\ref{fig6}.

\section{SUMMARY}
\label{summary}

With RV orbits for the last eight short-period binaries,
this last paper of the DDO series brings the number of the systems
studied at the David Dunlap Observatory to 141. There have
been 138 orbits contained in 14 of 15 papers of this series
(the seventh paper \citep{ddo7} did not contain any new data)
plus the separate investigations of W~Crv \citep{RL2000},
AW~UMa \citep{prib2008} and GSC~1387~475 \citep{rucpri2008}.

This paper -- in addition to the eight systems with the
reliable orbits -- summarizes all
remaining, unpublished spectroscopic material obtained
within the close-binary project. The included material
consists of systems with insufficient
phase coverage, problematic systems (too faint, too
southern for DDO or too complex multiples),
and systems found or suspected to be pulsating variables.
For most of the binaries of this paper,
we are presenting the first spectroscopic observations.
The highlights of the current study are:
(1)~the discovery of four triple systems
GO~Cyg, V563~Lyr, DD~Mon, V353~Peg;
(2)~the discovery of an under-luminous,
early-type component in V857~Her;
(3)~detection of prograde-rotating ripples
in BFs of CU~CVn and V752~Mon indicating non-radial pulsations.

Numerous discoveries and reliable solutions
of triple and quadruple systems show that the
BF de-convolution approach utilizing the SVD
method is a powerful and reliable technique which can be
applied not only to close binaries but also to other
types of spectral line broadening due to geometrical
effects. A good example of the power of this technique is our
detection of non-radial oscillations in two pulsating variables,
CU~CVn and V752~Mon. As far as we know,
the non-radial pulsations have been detected only in the
variability of line profiles of selected spectral lines
\citep{zima2006} or in cross-correlation (CCF) studies; the latter
obviously with a loss of information because
the CCF technique reduces the effective spectral resolution.
The only time the BF technique has been applied to
a $\delta$~Sct star (EE~Cam: \citet{breger2007})
indicated a potential for discovery of fine structure in line
profiles, probably akin to the ripples seen in CU~CVn and V752~Mon.

With the closure of the DDO observatory, this series has come
to an end; only a dozen or so known W~UMa-type eclipsing
binaries brighter than about $V = 10$ and potentially
accessible from DDO remain unobserved. Our
analysis for several stars presented here may feel
incomplete and unsatisfactory.
We still hope that the material presented will still lead to
or will supplement future follow-up observations of several binary
stars described here.

\acknowledgements

We express our thanks to Christopher Capobianco, George Conidis,
Yazya Ektiren, Kosmas Gazeas, Tomasz Kwiatkowski, Piotr Ligeza,
Stefan Mochnacki, Bogumil Pilecki, Wojtek Pych,
Matt Rock, and Piotr Rogoziecki for spectroscopic observations at DDO.

This study has been funded by the Canadian Space Agency Space Enhancement
Program (SSEP) with TP holding a Post-Doctoral Fellowship position at the
University of Toronto.
Support from the Natural Sciences and Engineering Council of
Canada to SMR and from the Polish Science Committee
(KBN grants PO3D~006~22 and
PO3D~003~24) to WO is acknowledged with gratitude.
The travel of TP to Canada has
been supported by a Slovak Academy of Sciences VEGA grant 2/7010/7.

The research made use of the SIMBAD database, operated at the CDS,
Strasbourg, France and accessible through the Canadian
Astronomy Data Centre, which is operated by the Herzberg Institute of
Astrophysics, National Research Council of Canada.
This research made also use of the Washington Double Star (WDS)
Catalog maintained at the U.S. Naval Observatory.

\clearpage

\noindent
Captions to figures:

\bigskip

\figcaption[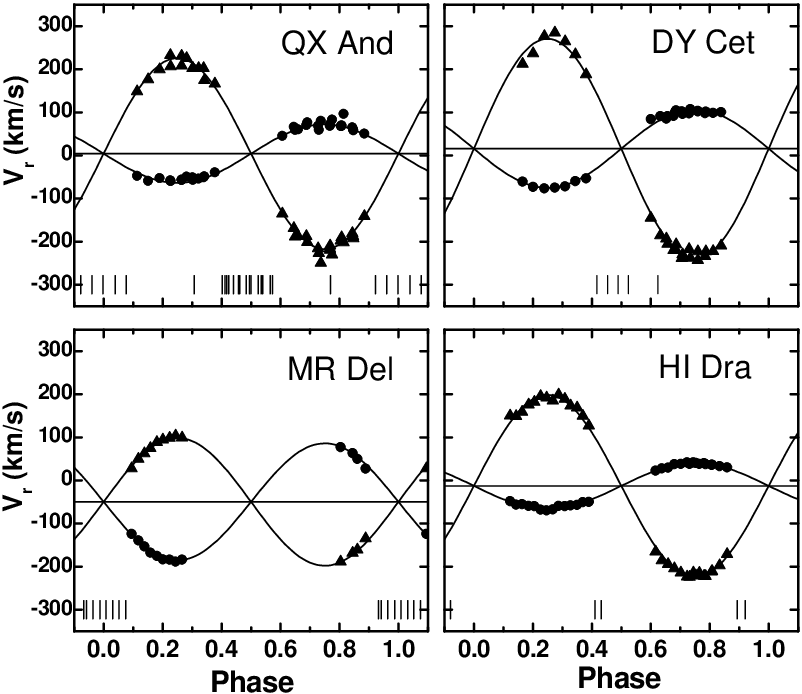] {\label{fig1} Radial velocities of four systems
with  spectroscopic orbits, QX~And, DY~Cet, MR~Del, HI~Dra,
are plotted in individual panels versus the orbital phases. The lines
give the respective circular-orbit (sine-curve) fits to the RVs.
MR~Del is triple system consisting of a close binary
and a slowly rotating single star. All systems, except MR~Del, are
contact binaries. The circles and triangles correspond
to components with velocities $V_1$ and $V_2$, as listed in
Table~\ref{tab1}, respectively. The component eclipsed at the minimum
corresponding to $T_0$ (as given in Table~\ref{tab2}) is the one which
shows negative velocities for the phase interval $0.0 - 0.5$ and
which is the more massive one. Short marks in the lower parts of the
panels show phases of available observations which were not used in
the solutions because of the spectral line blending or poor quality of data.
}

\figcaption[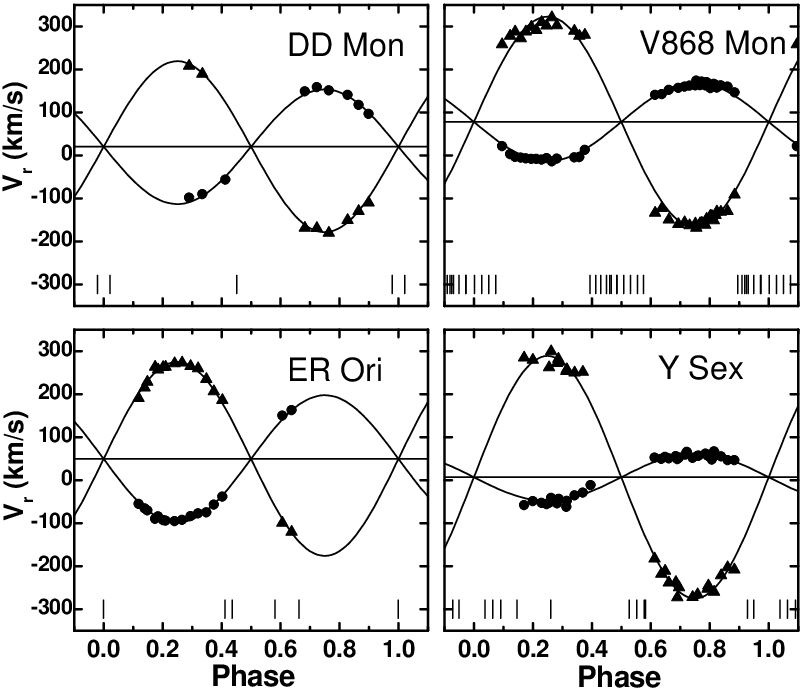] {\label{fig2} Radial velocities for the second
group of systems with  spectroscopic orbits, DD~Mon, V868~Mon,
ER~Ori, and Y~Sex. While V868~Mon is a contact binary, the remaining
systems are triple ones containing a contact or a close binary.
}

\figcaption[bfplots.eps] {\label{fig3}
The broadening functions (BFs) for all systems
where RV could be measured, as discussed in Sections~\ref{sec2} and
\ref{sec3}, in the constellation order.
The BFs were selected for orbital phases close to 0.25 or 0.75. The phases are
marked by numbers in the individual panels. Additional components to the
close binaries, MR~Del, V563~Lyr, DD~Mon, ER~Ori, Y~Sex, and LV~Vir are strong and
clearly visible. All panels have the same horizontal range, $-500$ to
$+500$ km~s$^{-1}$.
}

\figcaption[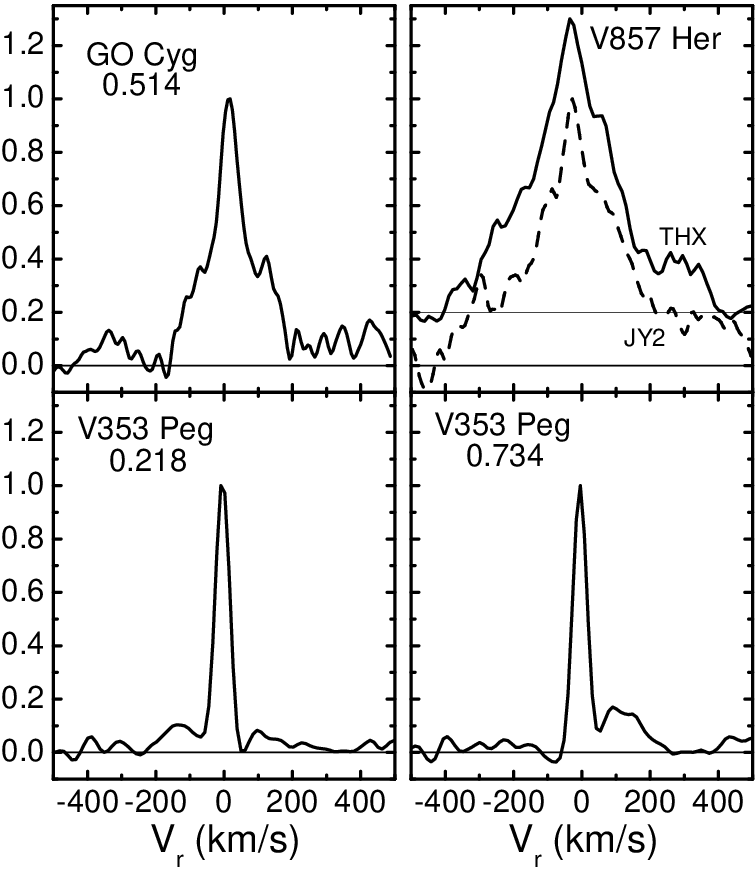] {\label{fig4}
The broadening functions (BFs) for problematic systems GO~Cyg,
V857~Her, and V353~Peg. For V857~Her broadening functions determined from
average spectrum obtained either using the 2160 lines/mm grating
and the JY2 chip or the 1800 lines/mm grating and the THX chip
are shown. In the case of V353~Peg, the presence of the binary
is demonstrated by the phase changes of its faint signature
between the two orbital quadratures.
}

\figcaption[bfproblem.eps] {\label{fig5}
Grayscale plots for two stars, CU~CVn and V752~Mon, very
apparently containing pulsating components.
The diagonally oriented ``ripples'' in both cases move
in the direction of the rotational motion. The BFs are sorted in time;
for V752~Mon, this applies to individual nights which are separated
by thin horizontal lines. One bin in the case of CU~CVn corresponds to about
7 minutes, while for V752~Mon we used longer exposures and one bin corresponds
to 17-18 minutes.
}

\figcaption[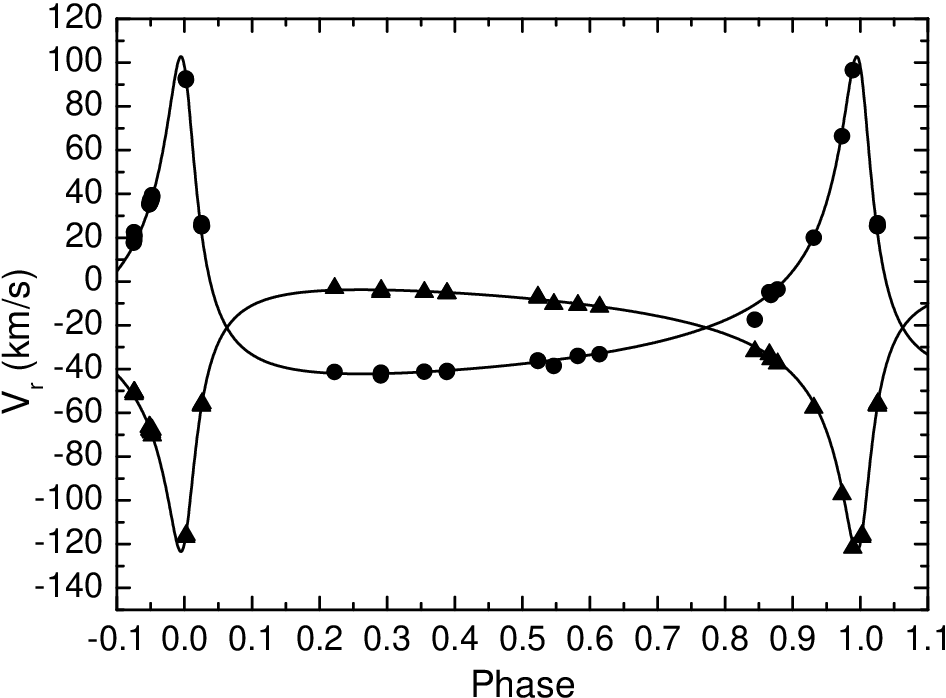] {\label{fig6}
Radial velocities of the 3rd and 4th components of the quadruple system
TV~UMi. The corresponding spectroscopic orbital solutions
are shown by continuous lines.
}

\clearpage

\addtocounter{figure}{-6}

\begin{figure}
\epsscale{0.85}
\plotone{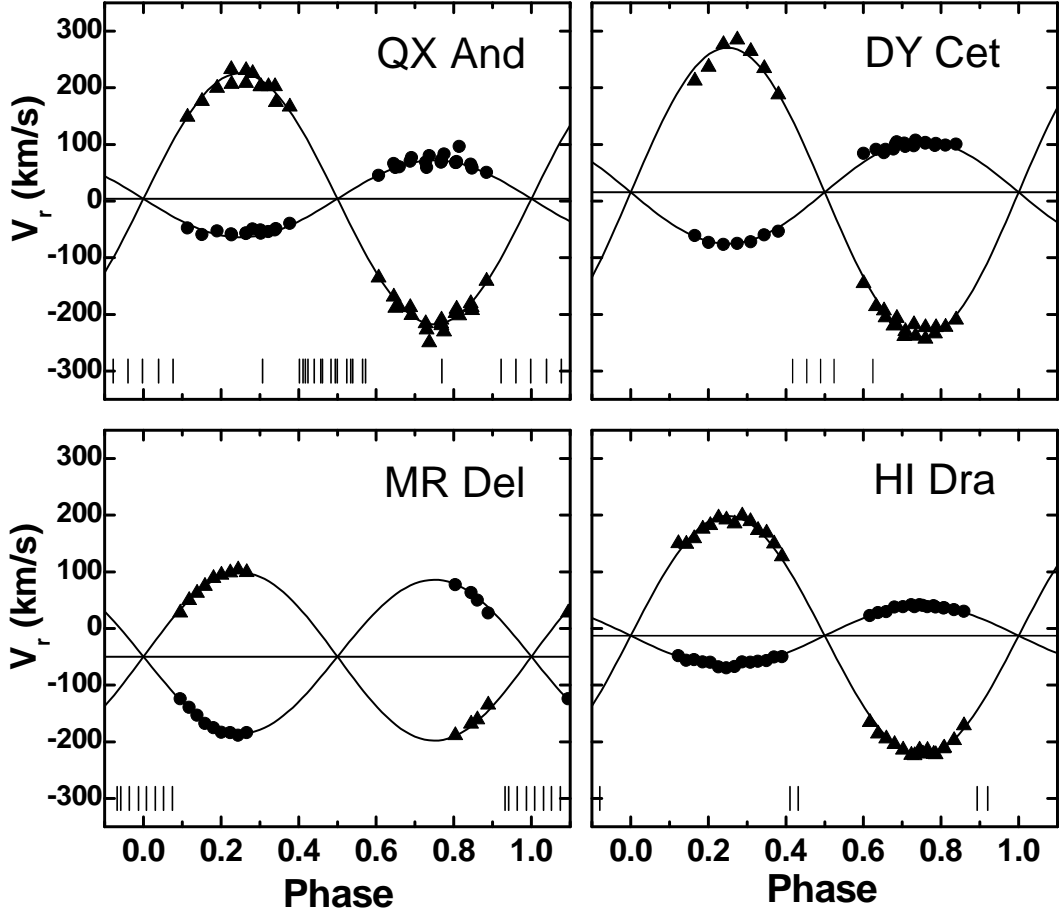}
\caption{Radial velocities of four systems
with  spectroscopic orbits, QX~And, DY~Cet, MR~Del, HI~Dra,
are plotted in individual panels versus the orbital phases. The lines
give the respective circular-orbit (sine-curve) fits to the RVs.
MR~Del is triple system consisting of a close binary
and a slowly rotating single star. All systems, except MR~Del, are
contact binaries. The circles and triangles correspond
to components with velocities $V_1$ and $V_2$, as listed in
Table~\ref{tab1}, respectively. The component eclipsed at the minimum
corresponding to $T_0$ (as given in Table~\ref{tab2}) is the one which
shows negative velocities for the phase interval $0.0 - 0.5$ and
which is the more massive one. Short marks in the lower parts of the
panels show phases of available observations which were not used in
the solutions because of the spectral line blending or poor quality of data.
}
\end{figure}

\begin{figure}
\epsscale{0.85}
\plotone{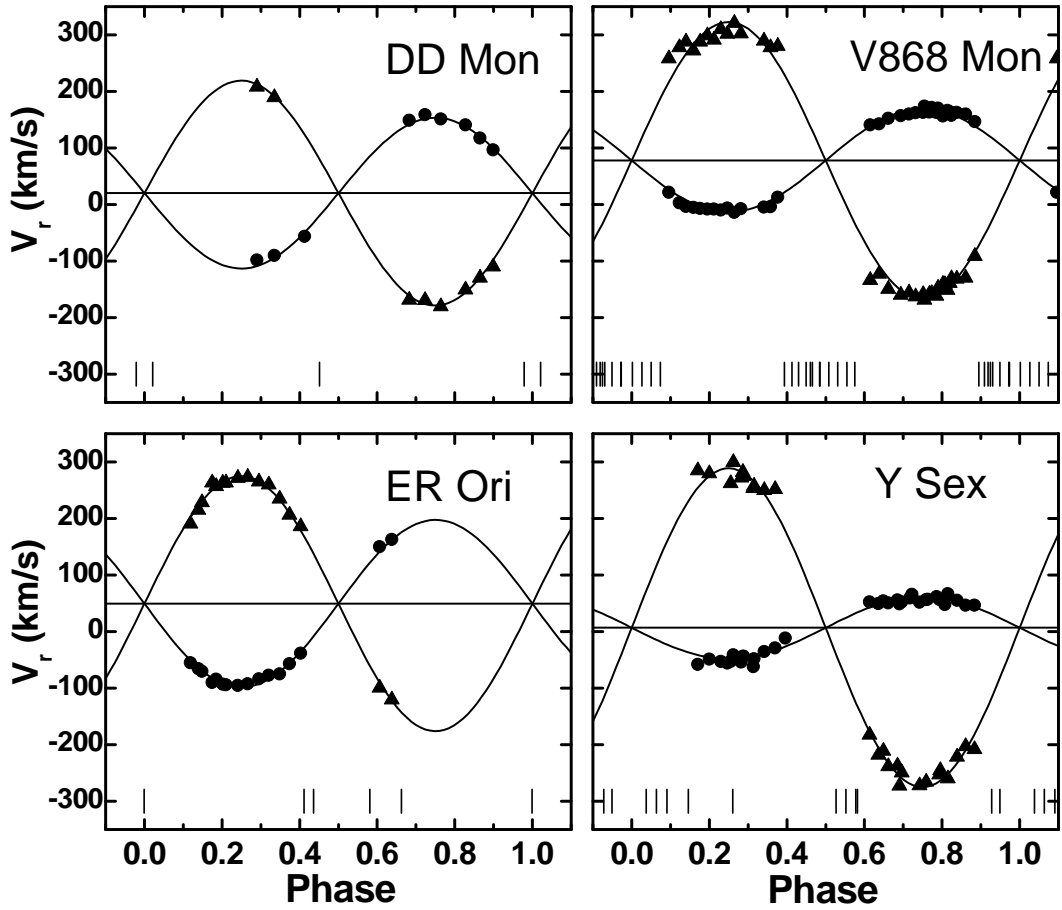}
\caption{Radial velocities for the second
group of systems with  spectroscopic orbits, DD~Mon, V868~Mon,
ER~Ori, and Y~Sex. While V868~Mon is a contact binary, the remaining
systems are triple ones containing a contact or a close binary.
}
\end{figure}

\begin{figure}
\epsscale{0.95}
\plotone{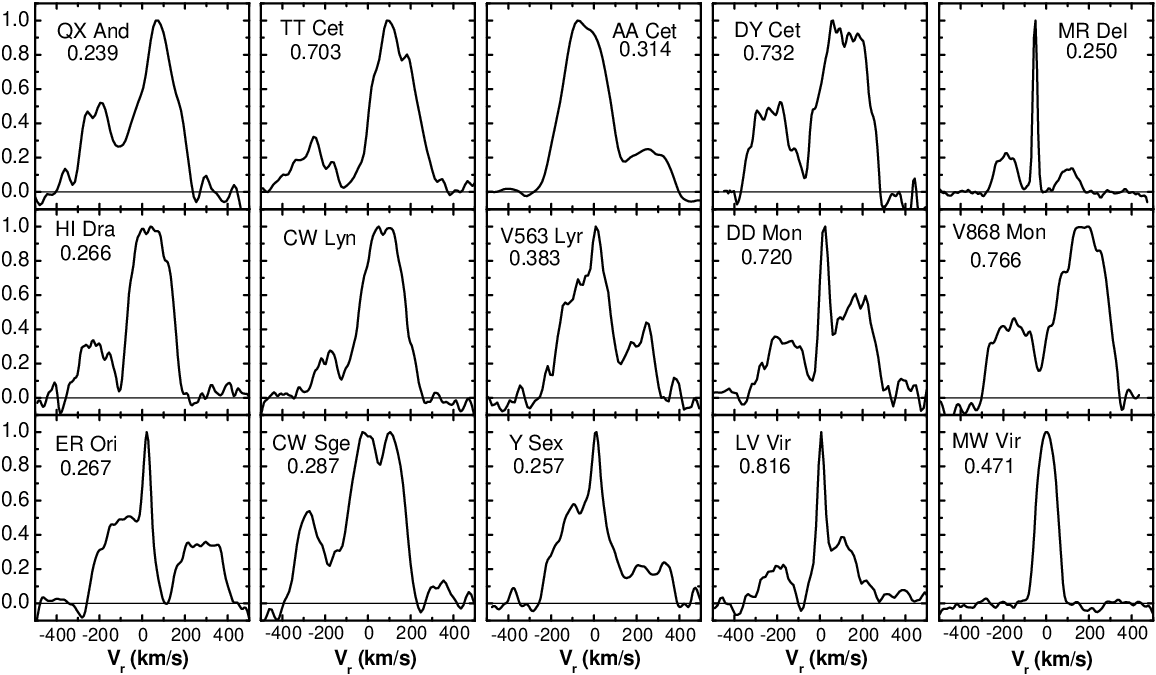}
\caption{The broadening functions (BFs) for all systems
where RV could be measured, as discussed in Sections~\ref{sec2} and
\ref{sec3}, in the constellation order.
The BFs were selected for orbital phases close to 0.25 or 0.75. The phases are
marked by numbers in the individual panels. Additional components to the
close binaries, MR~Del, V563~Lyr, DD~Mon, ER~Ori, Y~Sex, and LV~Vir are strong and
clearly visible. All panels have the same horizontal range, $-500$ to
$+500$ km~s$^{-1}$.
}
\end{figure}

\begin{figure}
\epsscale{0.75}
\plotone{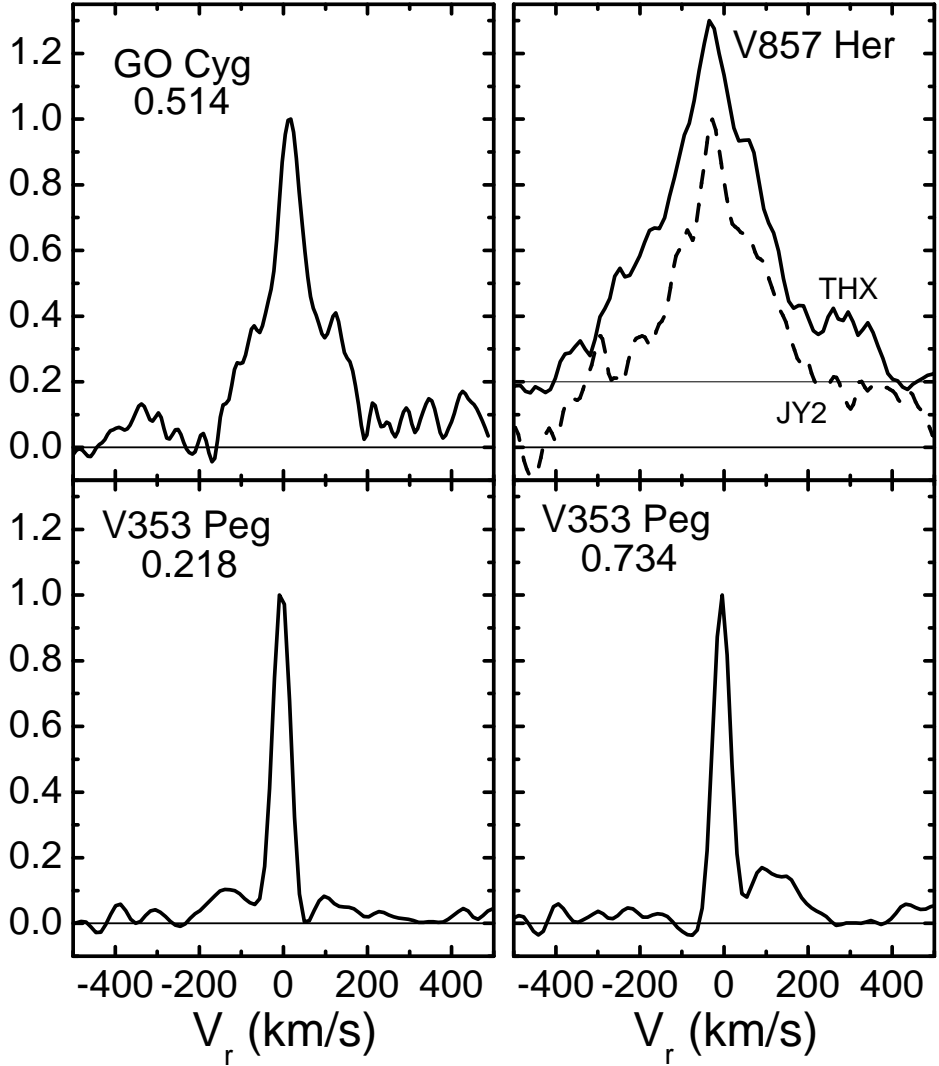}
\caption{The broadening functions (BFs) for problematic systems GO~Cyg,
V857~Her, and V353~Peg. For V857~Her broadening functions determined from
average spectrum obtained either using the 2160 lines/mm grating
and the JY2 chip or the 1800 lines/mm grating and the THX chip
are shown. In the case of V353~Peg, the presence of the binary
is demonstrated by the phase changes of its faint signature
between the two orbital quadratures.
}
\end{figure}

\begin{figure}
\epsscale{0.75}
\plotone{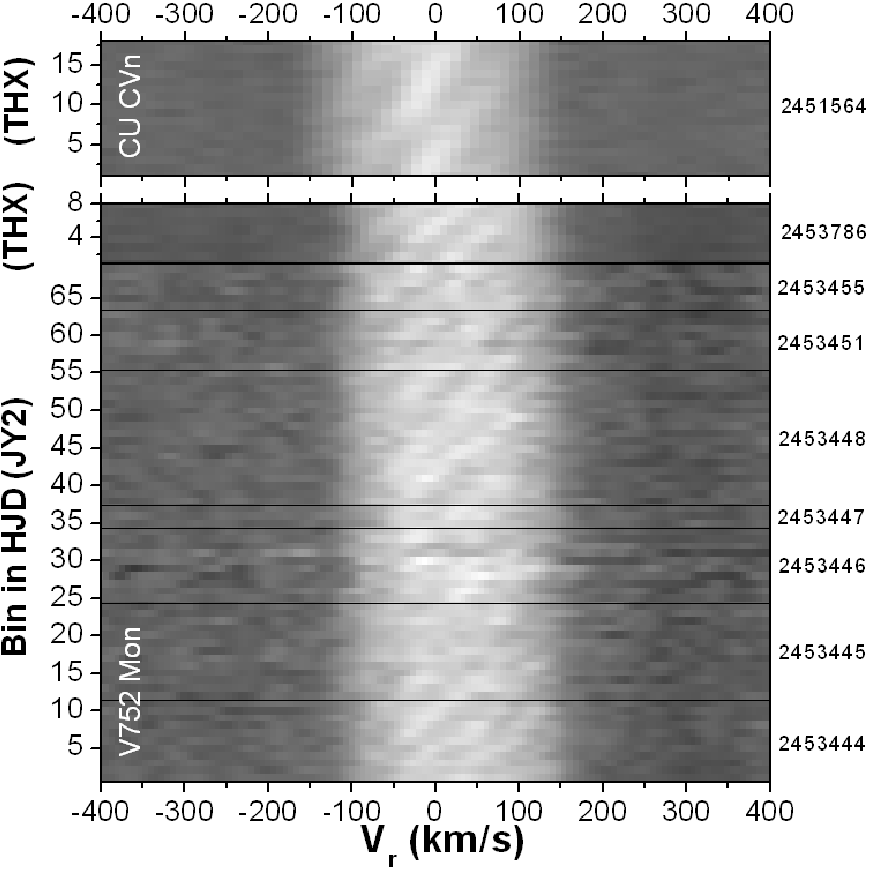}
\caption{Grayscale plots for two stars, CU~CVn and V752~Mon, very
apparently containing pulsating components.
The diagonally oriented ``ripples'' in both cases move
in the direction of the rotational motion. The BFs are sorted in time;
for V752~Mon, this applies to individual nights which are separated
by thin horizontal lines. One bin in the case of CU~CVn corresponds to about
7 minutes, while for V752~Mon we used longer exposures and one bin corresponds
to 17-18 minutes.
}
\end{figure}

\begin{figure}
\epsscale{0.75}
\plotone{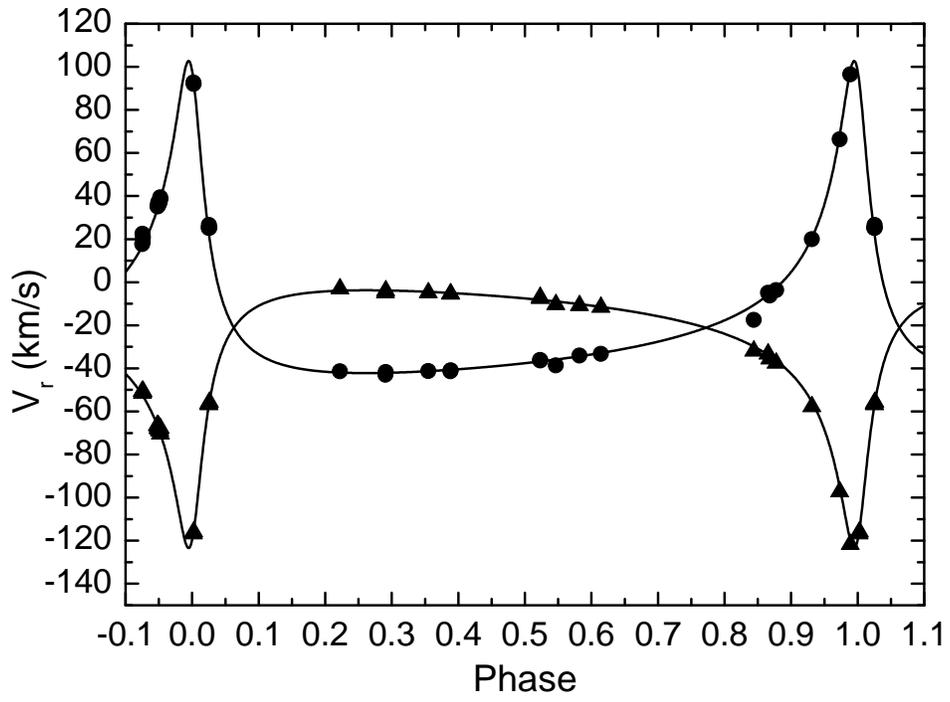}
\caption{Radial velocities of the 3rd and 4th components of the quadruple system
TV~UMi. The corresponding spectroscopic orbital solutions
are shown by continuous lines.
}
\end{figure}

\begin{deluxetable}{llrrrrr}

\tabletypesize{\footnotesize}

\tablewidth{0pt}
\tablenum{1}

\tablecaption{Radial velocity observations of close
binary stars
(the full table is available only in the electronic form)
\label{tab1}}

\tablehead{
\colhead{Target}          &
\colhead{HJD--2,400,000}  &
\colhead{~$V_1$}          &
\colhead{~~$W_1$}         &
\colhead{~$V_2$}          &
\colhead{~~$W_2$}         &
\colhead{Phase}          \\
                          &
                          &
\colhead{[km s$^{-1}$]}   &
                          &
\colhead{[km s$^{-1}$]}   &
                          & \\
}
\startdata
QX~And & 51433.6619 &$-49.23$ & 1.00 &   225.82  & 1.00 & 0.2814 \\
QX~And & 51433.6788 &$-54.06$ & 1.00 &   202.40  & 1.00 & 0.3224 \\
QX~And & 51433.7116 &   0.00  & 0.00 &     0.00  & 0.00 & 0.4020 \\
QX~And & 51433.7274 &   0.00  & 0.00 &     0.00  & 0.00 & 0.4403 \\
QX~And & 51433.7453 &   0.00  & 0.00 &     0.00  & 0.00 & 0.4837 \\
QX~And & 51433.7620 &   0.00  & 0.00 &     0.00  & 0.00 & 0.5242 \\
QX~And & 51433.7788 &   0.00  & 0.00 &     0.00  & 0.00 & 0.5650 \\
QX~And & 51433.7958 &  45.05  & 1.00 & $-135.18$ & 1.00 & 0.6062 \\
QX~And & 51433.8121 &  66.23  & 1.00 & $-168.90$ & 1.00 & 0.6458 \\
QX~And & 51433.8294 &  70.40  & 1.00 & $-187.34$ & 1.00 & 0.6878 \\
\enddata

\tablecomments{The table gives the RVs $V_i$ for observations of
close binary stars described in Sections~\ref{sec2} and \ref{sec3}
of the paper, in the order of constellation.
The first 10 rows of the table for a typical
program star, QX~And, are shown. Observations leading to entirely
inseparable broadening function peaks are given zero weight; these
observations may be eventually used in more
extensive modeling of broadening functions. Zero weights were assigned to
observations of marginally visible peaks of the secondary (sometimes
even primary) component. The RVs designated as $V_1$ correspond to
the more massive component; it was always the component eclipsed
during the minimum at the epoch $T_0$ (this does not always correspond
to the deeper minimum and photometric phase 0.0). The phases
correspond to spectroscopic $T_0$ given in Table~\ref{tab2} with the
exception of the following cases taken either from the
Hipparcos or \citep{Kreiner2004} databases: \\
TT~Cet: $52500.1309 + 0.4859541 \times E$ \\
AA~Cet: $52500.3652 + 0.5361685 \times E$ \\
CW~Lyn: $48500.457 + 0.812389 \times E$   \\
V563~Lyr: $52500.273 + 0.5776407 \times E$, \\
CW~Sge: $52500.567 + 0.6603633 \times E$, \\
LV~Vir: $52500.0315 + 0.4094453 \times E$, \\
MW~Vir: $48500.3032 + 0.493078 \times E$
(Hipparcos $T_0+P/2$ and $2\times P$).
}
\end{deluxetable}

\begin{deluxetable}{lccrrrccc}

\tabletypesize{\scriptsize}

\pagestyle{empty}
\tablecolumns{9}

\tablewidth{0pt}

\tablenum{2}
\tablecaption{Spectroscopic orbital elements for eight
binaries described in Sec.~\ref{sec2} \label{tab2}}
\tablehead{
   \colhead{Name} &                
   \colhead{Type} &                
   \colhead{Other names} &         
   \colhead{$V_0$~~~} &            
   \colhead{$K_1$~~~} &            
   \colhead{$\epsilon_1$~} &       
   \colhead{T$_0$ -- 2,400,000} &  
   \colhead{P (days)} &            
   \colhead{$q$}          \\       
   \colhead{}     &                
   \colhead{Sp.~type}    &         
   \colhead{}      &               
   \colhead{} &                    
   \colhead{$K_2$~~~} &            
   \colhead{$\epsilon_2$~} &       
   \colhead{$(O-C)$(d)~[E]} &      
   \colhead{$(M_1+M_2) \sin^3i$} & 
   \colhead{}                      
}
\startdata
QX~And     & EW(W?)              &             & $+4.07$(1.17)    & 67.92(1.82)  &
      8.18 & 51442.2015(11)      & 0.4121701   & 0.306(9)         \\ 
           & (F8)                &             &                  &221.73(1.84)  &
     11.39 & $+0.0045$~[$-$2,566.5]& 1.038(22) &                  \\[1mm]
DY~Cet     & EW(A)               & HD16515     & $+15.55$(1.49)   & 90.84(2.11)  &
      6.04 & 52594.3317(12)      & 0.440790(6) & 0.356(9)         \\ 
           & (F5V)               & BD$-14$ 495 &                  &255.40(2.28)  &
     12.06 & $-0.0005$~[+214.0]  & 1.896(40)   &                  \\[1mm]
MR~Del     & EB                  & HD195434    & $-49.76$(0.79)   &135.74(1.27)  &
      3.74 & 54469.1682(8)       & 0.5216899   & 0.915(12)        \\ 
           & (K2V+K6V?)          & BD+04 4470  &                  &148.32(1.27)  &
      4.35 & $+0.0015$~[+3,774.0]& 1.239(18)   &                  \\[1mm]
HI~Dra     & EW(A)               & HD171848    & $-12.73$(0.69)   & 52.83(1.08)  &
      2.30 & 54622.3721(10)      & 0.597418    & 0.250(5)         \\ 
           & (F0-2)              & BD+58 1824  &                  &211.49(1.08)  &
      7.43 & $+0.0228$~[+10,247.5]& 1.143(16)  &                  \\[1mm]
DD~Mon     & EB                  & HD292319    & $+20.21$(1.87)   &133.32(3.06)  &
      7.97 & 52642.7453(21)      & 0.568020    & 0.670(19)        \\ 
           & (G0)                &             &                  &198.85(3.18)  &
      8.87 & $-0.0022$~[+251.0]  & 2.157(65)   &                  \\[1mm]
V868~Mon   & EW                  & BD$-02$ 2221& $+78.08$(1.30)   & 91.21(1.83)  &
      6.46 & 54157.0662(18)      & 0.637705    & 0.373(8)         \\ 
           & (A5)                &             &                  &244.67(2.22)  &
     14.81 & $-0.0012$~[+1,724.0]& 2.504(52)   &                  \\[1mm]
ER~Ori     & EW(W)               & BD$-08$ 1050& $+49.45$(1.65)   &147.98(2.30)  &
      5.50 & 54393.4271(7)       & 0.4234034   & 0.656(12)        \\ 
           & F7/8                & HIP24156    &                  &225.48(2.35)  &
      6.93 & $+0.0099$~[+4,471.5]& 2.285(45)   &                  \\[1mm]
Y~Sex      & EW(A)               & HD87079     & $+6.78$(1.64)    & 54.96(2.30)  &
      7.69 & 54315.4217(13)      & 0.4198199   & 0.195(8)         \\ 
           & F5/6                & BD+01 2394  &                  &281.94(2.86)  &
     18.63 & $+0.0013$~[+4,324.0]& 1.663(45)   &                  \\
\enddata
\tablecomments{The spectral types given in column 2
relate to the combined spectral
type of all components in a system; they are given
in parentheses if taken from the
literature, otherwise they are new. The convention of naming
the binary components in the  table is that the
more massive star is marked by the subscript ``1'', so that the
mass ratio is defined to be always $q \le 1$.
The standard errors of the circular solutions in
the table are expressed in units of last decimal
places quoted; they  are given in parentheses
after each value. The center-of-mass velocities ($V_0$),
the velocity amplitudes ($K_i$) and the standard
unit-weight errors of the solutions
($\epsilon$) are all expressed in km~s$^{-1}$.
The spectroscopically determined  moments of primary
or secondary minima are given by $T_0$ (correspond
approximately to the average Julian date of the run);
the corresponding $(O-C)$ deviations (in days) have been
calculated from the available prediction on
primary minimum, as given in the text, using
the assumed periods and the number of
epochs given by [E].
The values of $(M_1+M_2)\sin^3 i$ are in the solar mass units.\\
Ephemerides ($HJD_{min}$ -- 2,400,000 + period in days)
used for the computation of the $(O-C)$ residuals:\\
 QX~And:    52500.0316 + 0.4121701 \\  
 DY~Cet:    52500.0031 + 0.4407903 \\  
 MR~Del:    52500.309  + 0.5216899 \\  
 HI~Dra:    48500.3186 + 0.597417  \\  
 DD~Mon:    52500.1745 + 0.568020  \\  
 V868~Mon:  53057.664  + 0.637705  \\  
 ER~Ori:    52500.1689 + 0.4234034 \\  
 Y~Sex:     52500.1192 + 0.4198199 \\  
 }
\end{deluxetable}

\begin{deluxetable}{llrrc}

\tabletypesize{\footnotesize}

\tablewidth{0pt}
\tablenum{3}
\tablecolumns{5}

\tablecaption{Radial velocity observations of the third and fourth
components of multiple systems (the full
table is available only in electronic form) \label{tab3}}
\tablehead{
\colhead{Target} & \colhead{HJD--2,400,000} & \colhead{$V_3$} & \colhead{~V$_4$} & \\
           &              & \colhead{[km s$^{-1}$]}  & \colhead{[km s$^{-1}$]} & \\
}
\startdata
ET Boo &  53812.95453 & $-77.29$ &  48.18 \\
ET Boo &  53842.77197 &          &        \\
ET Boo &  53843.74628 & $-45.12$ &  17.01 \\
MR Del &  54630.82080 & $-49.87$ &        \\
MR Del &  54638.68821 & $-50.06$ &        \\
MR Del &  54638.69974 & $-50.28$ &        \\
MR Del &  54638.71214 & $-50.20$ &        \\
MR Del &  54638.72308 & $-52.02$ &        \\
MR Del &  54638.73503 & $-49.74$ &        \\
MR Del &  54638.74573 & $-52.78$ &        \\
\enddata

\tablecomments{The table gives the RVs $V_i$ for
the third and fourth components. The 10 rows of the table
for the quadruple system, ET~Boo, and the triple system MR~Del are shown.
Observations of quadruple systems leading to entirely inseparable
broadening function peaks of components of the second binary have
been omitted from the table and not used in computation of the orbits
(the heliocentric Julian dates are, however, given).
The first three RV observations of TV~UMi (at HJD~2\,452\,694) were
found to be incorrect in the original publication \citep{ddo11}.
Correct values for RV of 3rd and 4th components are given here.}
\end{deluxetable}

\begin{deluxetable}{lrr}

\tabletypesize{\footnotesize}

\tablewidth{0pt}
\tablenum{4}

\tablecaption{Updated spectroscopic orbital elements of the second
(non-eclipsing) binaries in the quadruple systems ET~Boo, XY~Leo, and
TV~UMi \label{tab4}.
}
\tablehead{
\colhead{Parameter} &  & \colhead{error} \\
}
\startdata
\sidehead{\bf XY~Leo} 
$P_{34}$ [days]          & 0.8047497        &   0.0000042\\
$T_0$ [HJD]              & 2\,453\,882.1270 &   0.0008   \\
$V_0$ [km~s$^{-1}$]      & $-$39.97         &   0.22     \\
$K_3$ [km~s$^{-1}$]      &    46.44         &   0.34     \\
$f(m)$ [M$_\odot$]       &    0.00839       &   0.00018  \\
\sidehead{\bf ET~Boo} 
$P_{34}$ [days]          & 31.52135         &   0.00045  \\
$e_{34}$                 & 0.738            &   0.011    \\
$\omega$ [rad]           & 2.952            &   0.026    \\
$T_0$ [HJD]              & 2\,452\,930.737  &   0.027    \\
$V_0$ [km~s$^{-1}$]      &  $-24.09$        &   0.43     \\
$K_3$ [km~s$^{-1}$]      &    40.17         &   0.66     \\
$K_4$ [km~s$^{-1}$]      &    57.31         &   0.67     \\
$(M_3+M_4)\sin^3 i$ [M$_\odot$] & 0.928     & 0.052      \\
\sidehead{\bf TV~UMi} 
$P_{34}$ [days]          & 31.18836         &   0.00037  \\
$e_{34}$                 & 0.757            &   0.006    \\
$\omega$ [rad]           & 3.499            &   0.013    \\
$T_0$ [HJD]              & 2\,453\,192.994  &   0.016    \\
$V_0$ [km~s$^{-1}$]      &  $-21.12$        &   0.26     \\
$K_3$ [km~s$^{-1}$]      &    59.84         &   0.52     \\
$K_4$ [km~s$^{-1}$]      &    72.46         &   0.53     \\
$(M_3+M_4)\sin^3 i$ [M$_\odot$] & 2.084     &   0.063    \\
\enddata
\tablecomments{The table gives spectroscopic elements of the second
 binaries in the quadruple systems:
 orbital period ($P_{34}$), eccentricity ($e_{34}$),
 longitude of the periastron passage ($\omega$), time of the periastron
 passage ($T_0$), systemic velocity ($V_0$), semi-amplitudes of the
 RV changes ($K_3,K_4$). The corresponding mass ratio $q$, and total mass
 ($(M_3+M_4)\sin^3 i$) are given for ET~Boo, and TV~UMi where both
 components of the second binary could be measured.
 For the single-lined non-eclipsing binary in XY~Leo,
 only $f(m)$ is given. The orbit of the second binary in XY~Leo
 is circular, thus $e_{34} = 0.00$ and $\omega_{34} = \pi/2$.
 }
\end{deluxetable}

\begin{deluxetable}{lrrcrrlll}
\tabletypesize{\footnotesize}

\tablewidth{0pt}
\tablenum{5}

\tablecaption{Low-amplitude variable stars found to be
pulsating or of unknown type. \label{tab5}}
\tablehead{
\colhead{Target}       & 
\colhead{RV}           & 
\colhead{$v \sin i$}   & 
\colhead{Sp. type}     & 
\colhead{\# Obs.}      & 
\colhead{$\Delta H_p$} & 
\colhead{Period}       & 
\colhead{Class 1}      & 
\colhead{Class 2}     \\ 
                        & 
\colhead{[km~s$^{-1}$]} & 
\colhead{[km~s$^{-1}$]} & 
                        & 
                        & 
                        & 
\colhead{[days]}        & 
                        & 
                       \\ 
}
\startdata
FH~Cam    &   3.0  &  45   & A8    & 36/1 &  0.07  &  0.272478 &              &  puls. \\
CU~CVn    & $-4.1$ & 155   & A7    & 18/1 &  0.08  &  0.135667 &              &  puls. \\
V364~Cep  &  +8.9  & 153   & (A0)  & 10/1 &   -    &      -    &              &  puls. \\
V459~Cep  &  +1.2  &  97   & (F7)  &  3/1 &  0.05  &  0.178805 &  EW or puls. &  puls. \\
V2129~Cyg & $-22$  &  54   & (F8)  &  1/1 &  0.11  &  0.154876 &  puls.       &  EW or puls. \\
GW~Dra    &  var?  & $<$15 & (F2)  &  4/2 &  0.10  &  0.126184 &  puls.       &  puls. \\
IN~Dra    &  var?  & 147   & (F0)  &  5/3 &  0.06  &  0.137171 &  puls.       &  puls. \\
HV~Eri    & $-8.1$ &  83   & (F4)  &  1/1 &  0.14  &  0.210948 &  EW or puls. &  EW or puls. \\
PV~Gem    &  var?  & 135:  & F5-7  & 10/2 &  0.07  &  0.188065 &  EW or puls. &  puls. \\
V927~Her  & $-21$  & $<$15 & (F4)  &  1/1 &  0.17  &  0.130528 &  puls.       &  puls. \\
UX~LMi    &   var  &  60:  & (F7)  &  3/3 &  0.12  &  0.150638 &              &  puls. ? \\
CC~Lyn    &   var  &$<$15  & F0-2  & 62/4 &  0.10  &  0.354622 &  EW          &  puls + EW ? \\
V752~Mon  &   23   & 159   & (F0)  & 75/7 &  0.05  &  0.231451 &  EW          &  puls + EW ? \\
V1359~Ori & $-20.8$&  60   & (F7)  &  2/1 &  0.08  &  0.182158 &  EW or puls. & puls. ? \\
V579~Per  & $-3.5$ & 228   & A4/5  & 65/6 &  0.10  &  0.232812 &  EW          & puls.   \\
GG~UMa    & $-16.5$&  60   & (F6)  &  2/1 &  0.11  &  0.134841 &  puls.       & puls.   \\
GS~UMa    & $-3.4$ &  38   & (F5)  &  4/1 &  0.07  &  0.164007 &  puls.       & puls.   \\
\enddata
\tablecomments{The table gives the mean RVs for stars which are very
 probably pulsating variables or for multiple systems
 with one dominant variable component. Projected rotational velocities
 are given when $v \sin i > 30$ km~s$^{-1}$.
 Spectral types in parentheses have been taken from the literature, otherwise
 they are from our classification spectra. The extent of
 our observations can be judged from the column ``\#Obs'' where
 the number of observations is given
 as: No.\ of spectra/No.\ of nights.
 The Hipparcos amplitude, $\Delta H_p$
 (95th percentile -- 5th percentile) and the variability
 period are taken from the Hipparcos photometry Annex.
 For three systems,  FH~Cam, CU~CVn, and CC~Lyn, originally
 classified as $\beta$~Lyrae variables,
 this period corresponds to the double wave; for those stars, if
 definitely shown to be pulsating variables,
 one half of the Hipparcos period should be taken. The
last two columns ``Class 1''  and ``Class 2'', give the
variability classification following \citet{duer1997} and our
``best effort'' estimate, respectively.
 Stars known to be members of visual pairs in \citet{wds} are:
 HV~Eri (the components  about 30\arcsec apart),
 PV~Gem (the brighter component of a very wide visual
 pair, about $2\arcmin$ apart), CC~Lyn ($V = 6.62 + 8.26$, $\theta = 88\degr$,
 $\rho = 2.2\arcsec$), and V752~Mon ($V = 7.47+7.95$, $\theta = 24\degr$,
 $\rho = 1.7\arcsec$). The pulsating star
 V364~Cep was observed by us by mistake; it is the only
 variable listed in this table which was not
 discovered by the Hipparcos satellite mission.}
 \end{deluxetable}

\begin{deluxetable}{lccc}

\tabletypesize{\footnotesize}

\tablewidth{0pt}
\tablenum{6}
\tablecaption{Radial velocity observations of
single and pulsating stars (the full table is available
only in electronic form). \label{tab6}}
\tablehead{
\colhead{Target}         &
\colhead{HJD--2,400,000} &
\colhead{$RV$}           &
\colhead{$v \sin i$}     \\
                         &
                         &
\colhead{[km s$^{-1}$]}  &
\colhead{[km s$^{-1}$]}  \\
}
\startdata
FH Cam & 51658.65601 & 1.32 & 44.75 \\
FH Cam & 51658.71993 & 3.12 & 42.96 \\
FH Cam & 51658.72485 & 3.43 & 44.44 \\
FH Cam & 51658.72974 & 5.44 & 45.62 \\
FH Cam & 51658.73760 & 2.52 & 44.71 \\
FH Cam & 51658.74251 & 2.73 & 43.88 \\
FH Cam & 51658.74740 & 1.98 & 43.47 \\
FH Cam & 51658.75230 & 2.88 & 44.33 \\
FH Cam & 51658.76016 & 1.55 & 43.53 \\
FH Cam & 51658.76505 & 2.46 & 44.58 \\
\enddata
\tablecomments{Projected rotational velocity, $v \sin i$ is given only for
 rapidly rotating stars with $v \sin i > 30$ km~s$^{-1}$. For V752~Mon
 BFs determined from observations around 6290\AA~ lead to very scattered
 RVs - those are not given in the table}
\end{deluxetable}

\end{document}